\documentclass[twocolumn]{aastex63}

\usepackage{comment}
\usepackage[figure,figure*]{hypcap}
\usepackage{xcolor}

\usepackage{graphicx,natbib,placeins,amsmath}

\usepackage{multirow}
\usepackage{amssymb}
\usepackage{enumitem}
\usepackage{url}

\shorttitle{Companion detection and characterization through  KLIP}
\shortauthors{Strampelli et al.}

\begin{document}

\title{\textbf{\texttt{\textit{Stra}KLIP}}: A novel pipeline for detection and characterization of close-in faint companions through  Karhunen-Lo{\`e}ve Image Processing algorithm}

\correspondingauthor{Giovanni M. Strampelli}
\email{strampelligiovanni@jhu.edu}

\author[0000-0002-1652-420X]{Giovanni M. Strampelli}
\affiliation{Johns Hopkins University, 3400 N. Charles Street, Baltimore, MD 21218, USA}
\affiliation{Space Telescope Science Institute, 3700 San Martin Dr, Baltimore, MD 21218, USA}
\affiliation{Department of Astrophysics, University of La Laguna, Av. Astrofísico Francisco Sánchez, 38200 San Cristóbal de La Laguna, Tenerife, Canary Islands, Spain}

\author[0000-0003-3818-408X]{Laurent Pueyo}
\affiliation{Space Telescope Science Institute, 3700 San Martin Dr, Baltimore, MD 21218, USA}

\author[0000-0003-3184-0873]{Jonathan Aguilar}
\affiliation{Johns Hopkins University, 3400 N. Charles Street, Baltimore, MD 21218, USA}

\author[0000-0002-6054-0004]{Antonio Aparicio}
\affiliation{Department of Astrophysics, University of La Laguna, Av. Astrofísico Francisco Sánchez, 38200 San Cristóbal de La Laguna, Tenerife, Canary Islands, Spain}
\affiliation{Instituto de Astrofísica de Canarias, C. Vía Láctea, 38200, San Cristóbal de La Laguna, Tenerife, Canary Islands, Spain}

\author[0000-0002-5092-6464]{Gaspard Duchêne}
\affiliation{Astronomy Department, University of California, Berkeley, CA 94720, USA}
\affiliation{Université Grenoble Alpes/CNRS, Institut de Planétologie et d’Astrophysique de Grenoble, F-38000 Grenoble, France}

\author[0000-0002-9573-3199]{Massimo Robberto}
\affiliation{Johns Hopkins University, 3400 N. Charles Street, Baltimore, MD 21218, USA}
\affiliation{Space Telescope Science Institute, 3700 San Martin Dr, Baltimore, MD 21218, USA}


\begin{abstract}
We present a new 
pipeline developed 
to detect and characterize faint astronomical companions at small angular separation from the host star using sets of wide-field imaging observations not specifically designed for High Contrast Imaging analysis.
The core of the pipeline relies on Karhunen-Lo{\`e}ve truncated transformation of the reference PSF library to perform PSF subtraction and identify candidates. 
Tests of reliability of detections and characterization of companions are made through simulation of binaries and generation of Receiver Operating Characteristic curves for false positive/true positive analysis. The algorithm has been successfully tested on large \textit{HST}/ACS and WFC3 datasets acquired for two HST Treasury Programs on the Orion Nebula Cluster. Based on these extensive numerical experiments we find that, despite being based on methods designed for observations of single star at a time, our pipeline performs very well on mosaic space based data. In fact, we are able to detect brown dwarf-mass companions almost down to the planetary mass limit. The pipeline is able to reliably detect signals at separations as close as $\gtrsim 0.1 ''$ with a completeness of $\gtrsim 10\%$, or $\sim 0.2''$ with a completeness of $\sim 30\%$. This approach can potentially be applied to a wide variety of space based imaging surveys, starting with data in the existing HST archive, near-future JWST mosaics, and future wide-field Roman images.
\end{abstract}


\section{Introduction}
The ability to detect and analyse companions in binaries and multiple systems is essential for characterizing the frequency and physical properties of those objects. Linking their parameters to models of dynamical evolution from their original molecular cloud  can provide insight on the physics at play during the earliest  stages of star formation. For this purpose, the multiplicity of pre-main sequence stars has been investigated over a broad range of environments: dense stellar cluster \citep[e.g.][]{Petr1998,Kohler2006,Reipurth2007,Luhman2005}, young OB associations \citep[e.g.][]{Brown1997,Shatsky2002,Kouwenhoven2007}, T associations \cite[e.g. ][]{Kraus2008,Kraus2011}. Speckle interferometry, adaptive optics, aperture masking and 
high-contrast imaging (HCI, 
e.g. coronography and nulling interferometry)
are just some of the techniques that have been applied to obtain rich datasets. 

In principle HCI is a most promising technique. However, even when  the primary star can be masked out, 
the detection of faint close-in companions is heavily hampered by the dominating presence of bright quasi-static speckles mainly caused by imperfections in the optics \citep{Schneider2003,Biller2004,Marois2005,Masciadri2005}.
To deal with this problem, and possibly take advantage of the quasi-static nature of the speckles, a number of PSF modeling algorithms have been developed: LOCI \citep[Locally Optimized Combination of Images algorithm;][]{Lafrenire2007}, NNMF \citep[Non-negative Matrix factorization: Robust Extraction of Extended Structures;][]{Ren2018}, pynpoint \citep{Amara2012,Stolker2019}, and KLIP \citep[Karhunen-Lo{\`e}ve Image Processing algorithm;][]{Soummer2012}

In this paper we present an application of KLIP to detect faint astrophysical companions at small angular separation in regular wide-field images taken with the Hubble Space Telescope, i.e. observations not specifically designed for HCI analysis. We focus in particular on data taken with Wide Field Channel of ACS at visible wavelength and with the IR channel of WFC3, that can be considered representative of two different Point Spread Function sampling regimes. In Section $\S 2$ we present a general overview of the pipeline, while in Sections $\S 3$ to $\S 8$ we describe more in depth each individual steps. The conclusions are presented in $\S 9$. 
Our results are expressed using two metrics: completeness/robustness  of a detection using a ROC-based approach and photometric uncertainty on the measured brightness of discovered companion. In theory astrometric precision on a companion could be  considered but precise orbit fitting is usually carried out with narrower field observations and we thus omit it in the context of this paper.


\begin{figure*}
    \centering
    \includegraphics[width=0.95\textwidth]{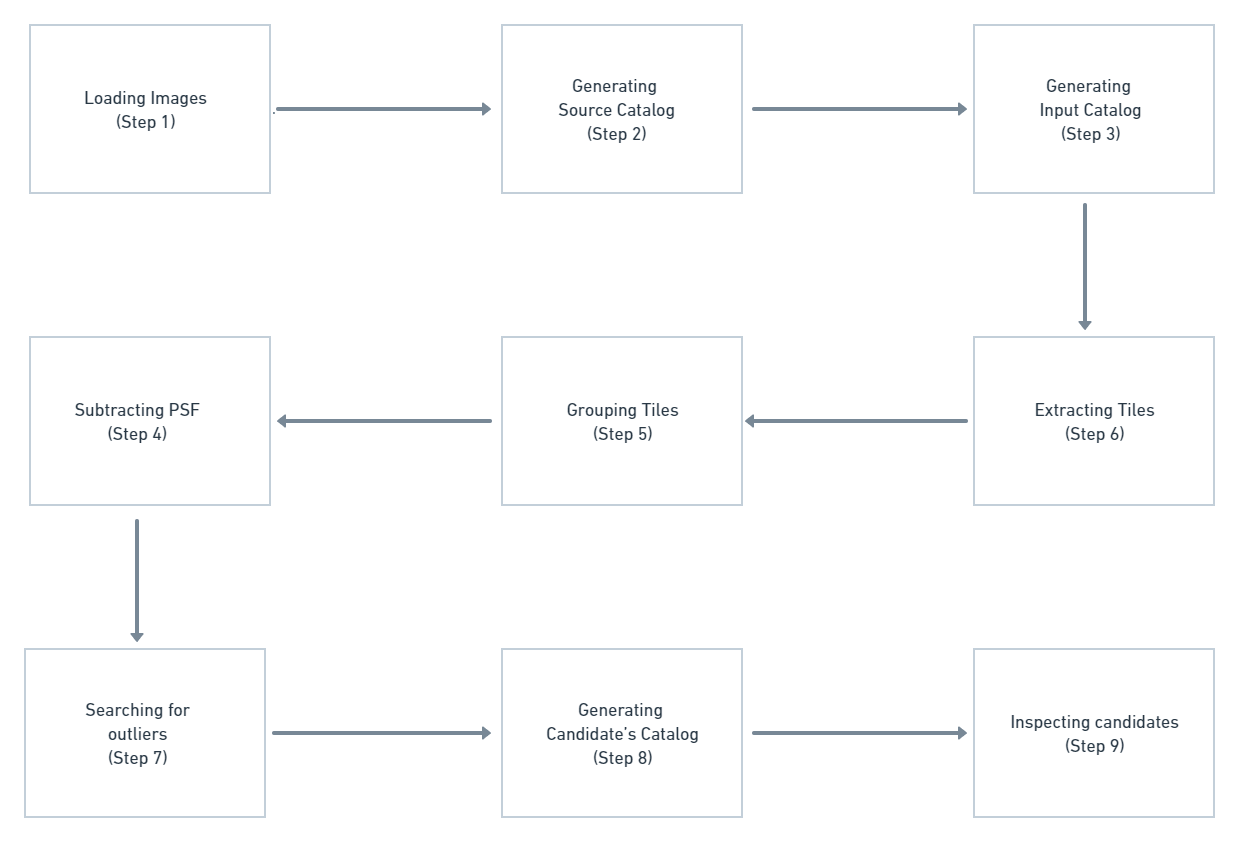}
    \caption{Visualization of the pipeline major steps.}
    \label{fig:/StraKLIP flowchart}
\end{figure*}

\section{Overview of the pipeline}
The core of the pipeline, coded in Python 3.7 \citep{Python}, can be divided in nine basic steps  that will be briefly introduced here and described more in detail later in the following sections.  
Due to the highly interactive nature of the analysis, the default version of pipeline (called  \textbf{\texttt{\textit{Stra}KLIP}}) organizes the different steps in a series of Jupyter notebooks\footnote{The Jupyter Notebook is an open-source web application that allows to create and share documents that contain live code, equations, visualizations and narrative text. \url{https://jupyter.org/index.html}} (from here after simply notebook) 

The main objective is to detect faint companions by perform PSF subtraction using the KLIP algorithm, Step \ref{item:point6}.
This requires two types of input: 
\begin{itemize}
    \item a set of \textit{initial \_flt HST} images, as delivered by the standard HST pipelines.
    \item an \textit{initial catalog} of primary targets to be searched for companions. Note that the existence of the initial catalog is a \textit{soft} requirement - see point \ref{item:point2} later in this Section).
\end{itemize}
During the following discussion, we will distinguish between the \textit{initial catalog} of target sources, typically the result of previous data analysis aimed at deriving a photometric source catalog,
and the derived \textit{input catalog} in the format actually accepted by the pipeline. A similar distinction applies to the images: the prefix \textit{initial} refers to the \textit{\_flt HST images} delivered by the standard pipeline and downloaded from the Space Telescope Science Institute (STScI) MAST archive; they are different from the \textit{input images} needed by the pipeline.
These are small \textit{FITS} images cut around each target source; their size define the searching region $\mathcal{S}$ for the companions. We shall refer to them as \textit{postage-tiles} or \textit{tiles}, interchangeably.
\vspace{5mm}

The nine steps of the pipeline are (see also Figure \ref{fig:/StraKLIP flowchart}):
\begin{enumerate}
    \item{\bf Loading images:} a preliminary step needed to adapt the \textit{initial images} to the format expected by the pipeline. The pipeline require a set of \textit{initial \_flt images} in electrons/sec, already multiplied by the pixel area map (PAM\footnote{\textit{HST}/WFC3-IR and UVIS:\url{https://www.stsci.edu/hst/instrumentation/wfc3/data-analysis/pixel-area-maps}}\footnote{ACS:\url{https://www.stsci.edu/hst/instrumentation/acs/data-analysis/pixel-area-maps}.}). This is an image where each pixel value describes the pixel's area on the sky relative to the native plate scale, and allow to account for differences between on-sky pixel size across the field of view (FOV) in images that have not been distortion corrected.
    For each filter, the pipeline rearranges the full set of images as a multi-dimensional \textit{datacube} with the following layers: 1) the \textit{SCI} labeled image that will contain the actual data, 2) the \textit{DQ} labeled image containing the data quality flags and 3) the \textit{ERR} labeled image containing the uncertainties associated to the \textit{SCI} image;\label{item:point1}
    
    \item{\bf Generating a source catalog:} If the \textit{initial catalog} only provides the positions of each source on each individual image, preliminary aperture photometry can be performed on all sources using the class \textbf{\texttt{photometry$_{AP}$}} (see later in Sec. \ref{sec:Aperture photometry});
    \label{item:point2}.
    If an \textit{initial catalog} is not available, a collection of routines has been developed to create it use \textbf{\texttt{TweakReg}} package from DrizzlePac\footnote{\url{https://www.stsci.edu/scientific-community/software/drizzlepac.html}} to align the \textit{HST} images to the Gaia reference catalog \citep{Gaia2018} (or any other catalog of sources provided).
    This gives the pipeline the versatility needed to work regardless of the accuracy, or even the existence of an \textit{initial catalog}. 
    
    \item{\bf Generating an input catalog:} In this step the \textit{initial  catalog} of target sources is modified and reassembled in order to produce an \textit{input catalog} for the pipeline in a suitable format;\label{item:point3}
    
    \item{\bf Extracting tiles:} Small \textit{tiles} are  created for each entry of the \textit{input catalog}, trimmed so that the centroid of each  star coincides with the center of its \textit{tile}, setting-up a search region $\mathcal{S}$ (see Section \ref{Sec:KLIP PSF subtraction and candidate detections}) for each target; \label{item:point4}
    
    \item{\bf Grouping tiles:} The \textit{tiles} are grouped according to their position in the FOV of the instrument to be processed to determine suitable field-dependent PSF stars;\label{item:point5}
    
    \item{\bf Subtracting PSFs:} PSF subtraction is performed on each \textit{tile} using the KLIP algorithm \citep{Soummer2012};\label{item:point6}
    
    \item{\bf Searching for residual outliers:} The residual \textit{tiles} produced by the previous step are analyzed for the presence of a previously undetected astronomical source;\label{item:point7}
    
    \item{\bf Generating a catalog of candidates:} An output catalog of candidate detections, with their photometry, is produced.\label{item:point8}
    
    \item{\bf Inspecting candidates:} A collection of tools is used to assess the reliability of the detections. They are based on extensive simulations where both false negative (missed injected companion) and false positive (noise residuals classified as companions) are statistically analyzed. 

\end{enumerate}

The final product of the pipeline is represented by  \textit{pandas}\footnote{\url{https://pandas.pydata.org/docs/index.html}} dataframe and stored in a hierarchical data format \citep[HDF5; ][]{McKinney2010}.
recording the output of the different steps of the pipeline.
A library of notebook allows to analyze the output of the pipeline and produce the type of results shown e.g. in \cite{Strampelli2020,Strampelli2022}. 
Again, the notebook format has been chosen because of the highly interactive nature of the final analysis. They provide examples of the ancillary routines developed to analyze and interpret the output catalog, providing warnings with the capability of rejectings candidates that do not pass the reliabilty threshold for false positive detections. 


\section{Step 1: Loading images}
The \textit{input images} accepted by the pipeline are the \textit{flt HST} images delivered by the STScI MAST archive. While each fits file is composed by five layers,  the pipeline considers only three of them:  
\begin{itemize}
    \item science image, extension \textit{SCI}, where the actual science data reside;
    \item data quality, extension \textit{DQ}, containing coded information about bad pixels, saturated pixels, cosmic rays etc.;
    \item error layer, extension \textit{ERR}, providing the variance of the science image.
\end{itemize}


The initial \textit{split\_chip.py} routine takes advantage of \textit{ACSmask/WFC3mask} from DOLPHOT package \citep{Dolphin2000}\footnote{See the packages handbook  at \url{http://americano.dolphinsim.com/dolphot/} for more details} to  mask out all pixels flagged as bad in the data quality image, so that they will be ignored in the future steps (sky determination, photometry, etc.) and multiplies the \textit{initial image} by the PAM.
While the  \textit{WFC3-IR} are \textit{single-chip} image datacubes,
in the case of  \textit{ACS-WFC} and \textit{WFC3-UVIS}, the routine distinguishes between the two CCDs of each camera splitting the \textit{initial images} in two \textit{.chipN} \textit{FITS} datacubes, where \textit{N} represent the chip number 1 or 2. Both datacubes contain their relative \textit{SCI}, \textit{DQ} and \textit{ERR} sub-layers. 
Before running this step it is best practice to backup all the \textit{initial images} because the \textit{ACSmask/WFC3mask} from DOLPHOT package will alter them. 

\section{Step 2 and 3: Catalogs}
\label{Sec:Step 2 and 3: Catalogs}
Assuming an \textit{initial catalog} has been created, the pipeline generates four \textit{pandas} dataframes, namely:
\begin{itemize}
    \item a \textit{header} dataframe storing all the global  information the pipeline will need in order to deliver the final results. These include project-dependent column labels, selected pipeline input options, information about the detector like pixel-scale, detector dimensions, etc. This dataframe provides also  a 'recovery mode' that keeps trace of the  options selected during the different steps of the pipeline, allowing the user to go back and restart from any intermediate step;
    
    \item a \textit{multiple visits} dataframes. A \textit{multiple visits target} dataframe  holds the original position and photometry of each source in the \textit{initial images} as provided by the \textit{initial catalog}, with their uncertainties. 
    Ancillary information is also included, such as the program \textit{visit}, the camera \textit{CCD}, the telescope position angle \textit{PA\_V3}, \textit{saturation}, \textit{image name} and more. Sources detected in different visits have one entry for each detection, each detection receiving a unique label. 
    
    
    \item similar to the previous \textit{multiple visits} dataframes, the pipeline creates a \textit{average} dataframe. The \textit{average target} dataframe holds the averaged photometry and celestial coordinates of the sources that are detected multiple times, along
    with their unique identifier: \textit{average ID}. 

    \item a \textit{cross-match} dataframe links together the IDs from the \textit{multiple visits} dataframe to the \textit{average} ones.
    
\end{itemize}

A \textit{type} keyword is added to the \textit{average target} dataframe to help during some crucial decision-making points along the pipeline (see Table \ref{tab:type}). Sources flagged as \textit{type 0} are rejected by the pipeline due to the possible poor photometry, noisy image, presence of spikes, cosmic rays, image artifacts or any other reasons that could bring to a wrong detection.
\textit{Type 1} instead are the typical target for the pipeline. They usually are isolated have a good photometry (within user provided parameters). 
\textit{Type~2} labels close pairs that can be resolved but have not been reliably measured in the \textit{initial catalog}  (see Figure \ref{fig:LACR} for an example of a \textit{type~2} source).
\textit{Type~3} sources are well resolved pairs with individual  photometry. The pipeline will ignore these targets unless told otherwise.

\begin{table}[!t]
    \centering
    \begin{tabular}{cl}
        \hline
        \textit{type} & Explanation\\
        \hline
        \hline
        0 &	a target rejected from/ the pipeline\\
        1 &	a good target for the pipeline\\
        2 &	unresolved double \\
        3 &	known double\\
        n & user defined flag\\
        \hline
    \end{tabular}
    \caption{\textit{type} flag entry adopted by the pipeline. Only source with a \textit{type 1} will be selected as possible PSF reference. Sources with \textit{type~1,2} or \textit{n} (where $n \not\in [0,1,2,3]$ ) will be processed by the pipeline in search for a companion while \textit{type~0}  or \textit{3} sources will be skipped.}
    \label{tab:type}
\end{table}

\textit{Type~2} flag is not added automatically by the pipeline, instead, it can be added manually through the \textit{Showroom} routine that allow the user to visually inspect each produced tile and modify their \textit{type}. 
\textit{Type~2} sources provide a crucial test to the overall performance of the pipeline. Tuning the many options available at each step of the pipeline one should  be  able to recover most - if not all - of these not-fully-resolved binary systems. On the other hand, resolved pairs pose a problem for the creation of a good PSF. For this reason, \textit{type 3} source are ignored as their fluxes can be derived using more conventional PSF photometry techniques. Higher types are used to flag non-stellar sources (e.g. galaxies, proplyds, etc.)


Entries in the \textit{multiple visits target} dataframe get a new flag to indicate if the source is either a \textit{good, unresolved or bad candidate} for PSF subtraction (corresponding to the previous \textit{type-1, 2 or 0}), a \textit{good PSF} star (also a \textit{type-1} but with user define limits on photometry) or a \textit{wide-double} (\textit{type-3}). In particular, to be selected as a PSF star, a source must be bright and well isolated, but not  saturated. This selection can be made parsing a maximum number of saturated pixel in the tile, as well as a user-defined range of magnitudes and errors. \textit{Good candidates} are selected within a user-defined range of magnitudes and errors. \textit{Bad candidate}, i.e. sources either saturated, not detected in the filter or too faint, as well as \textit{known doubles} (\textit{type 3}) are ignored. This means that any \textit{type} different from 0 an 3 can be treated by the pipeline.  The pipeline will still run on them and the residuals can easily be retrieved at the end calling the right  \textit{type} flag.

A last flag is added to the \textit{multiple visits target} dataframe to indicate the \textit{quadrant} (or \textit{cell}) where the source appears on the detector. \textit{Cell} have  equal size and should be large enough to contain enough PSF stars so that different \textit{KLIP-modes} can be investigated during the PSF subtraction step. This is because the number of \textit{KLIP-modes} chosen to truncate the  Karhunen-Lo{\`e}ve transformation is strictly related to number of reference stars in the \textit{PSF reference library}. On the other hand, the size of the \textit{cells} should be small enough to avoid spatial variations of the PSF, providing a local PSF model. The user can decide how many equal-size cell divide the field of view.

\begin{figure*}
    \centering
    \includegraphics[width=0.95\textwidth]{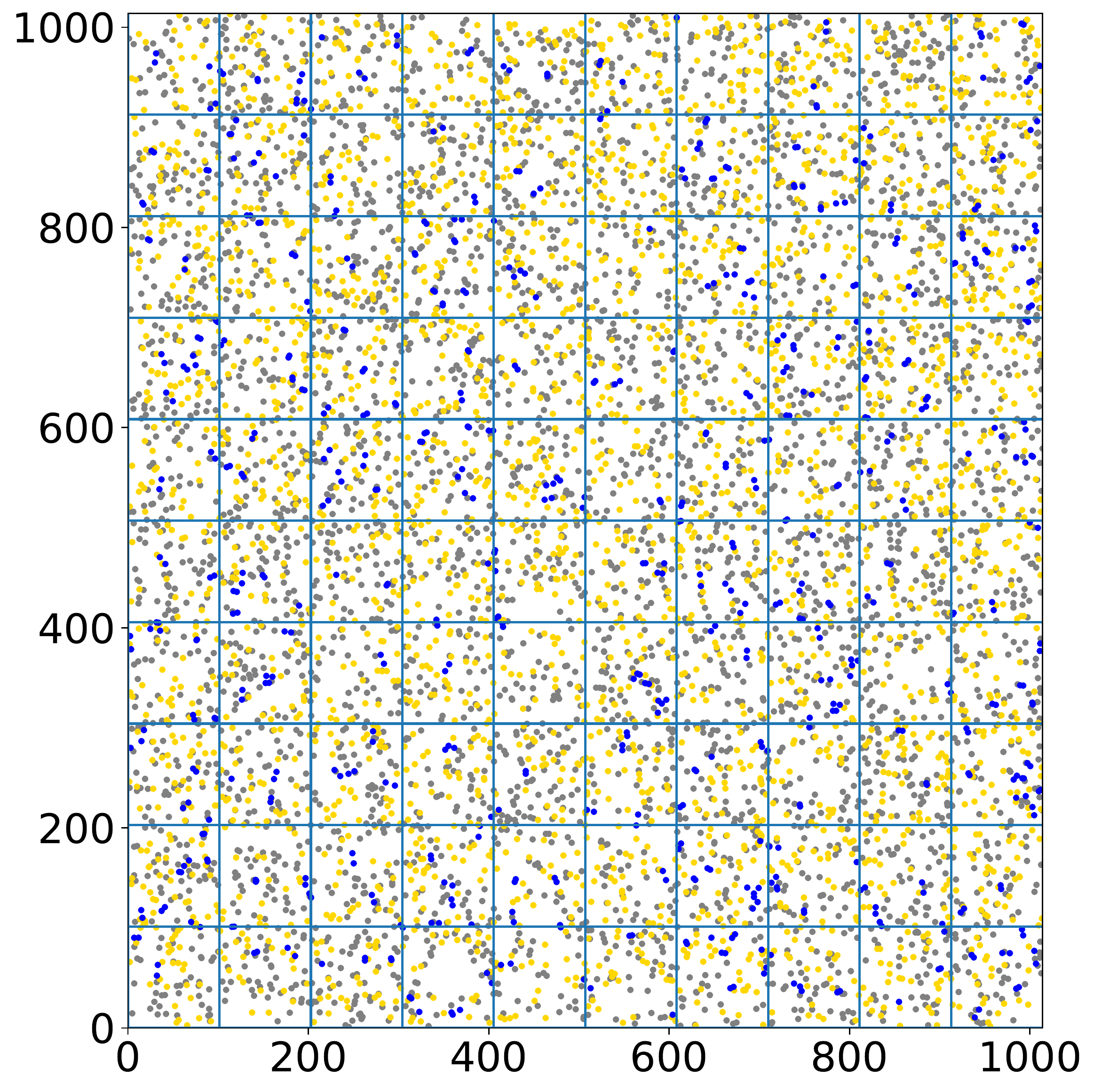}
    \caption{\textit{WFC3-IR} FOV (1014x1014 pixels). Gray dots mark the position on the instrument of selected \textit{good candidates}, while yellow dots mark \textit{PSF} stars. Blue dots mark the positions of identified \textit{wide doubles}. In this example we settled on one hundred \textit{cells} with an averaged number of isolated targets (\textit{good candidates}) of $\sim 44$ and averaged number of \textit{PSF} stars of $\sim 32$.}
    \label{fig:quadrant}
\end{figure*}

A practical example is provided in Figure \ref{fig:quadrant}, where the \textit{WFC3-IR} FOV has been divided in $10\times 10$ cells.  Good candidates (gray dots), PSF reference stars (yellow dots) and known doubles (blue dots) are shown on the plot. The particular choice for this partition of the focal plane allows to maintain the PSF asymmetry within a nominal $1\%$ variation, with about 50 sources per cell.


The pipeline saves an output dataframe as an HDF5 file, composed of the six \textit{header, average/multiple visits target} dataframes and the \textit{cross-match IDs} dataframes. It will be updated in the following step as the data relative to the target and residual tiles (as output of the PSF subtraction) and photometry are derived. 

\section{Steps 4 and 5: Tiles}
\label{sec:Making the tiles}
During this step the pipeline runs through every entry in the \textit{multiple visits target} dataframe that is not labeled as \textit{known double} and creates a \textit{tile} centered on the \textit{(x,y)} coordinates of the source on the \textit{input image}. The size of the tile can be defined by the  user, the default being 1.5 arcsec. 
In setting the dimension of the tile, one has to consider the following factors:
\begin{itemize}
    \item the area  must be large enough to contain the bright wings of the PSF, for sources matching our assumed range of magnitudes;
    \item the area must have enough pixels to provide a meaningful noise calculation. Detections of close companions are affected by small number statistics and a correction to the  estimated contrast and SNR has to be applied \citep{Mawet2014}. 
    \item the area must be small enough so that tiles do not overlap. At the actual state of development, the pipeline the routine are able to detect only the first brighter companion in a multiple system, so if the companion is already present in the \textit{input} catalog, the pipeline will skip this system (see type 3 sources earlier in Section \ref{Sec:Step 2 and 3: Catalogs});
\end{itemize}

To create a tile for analysis, the routine first creates an initial version using one of the following options, before cleaning cosmic rays and refining the centroid: 
\begin{itemize}
\item \textit{input coordinates}: use the coordinate from the \textit{input catalog};
\item \textit{reference filter}: use the coordinates from the filter with the smallest magnitude error, excluding detections labeled as \textit{bad}. This option should only be used when  the exposures in different filters are taken back to back, without changing the telescope pointing, and the source is barely detected in some of the filters;
\item \textit{reference ID}: use the average coordinates of all detections in the \textit{same} filter. Again,this may be handy when the pointings between two visits is the same and the sources are faint.
\end{itemize}

\begin{figure*}
\centering
\includegraphics[width=0.45\textwidth]{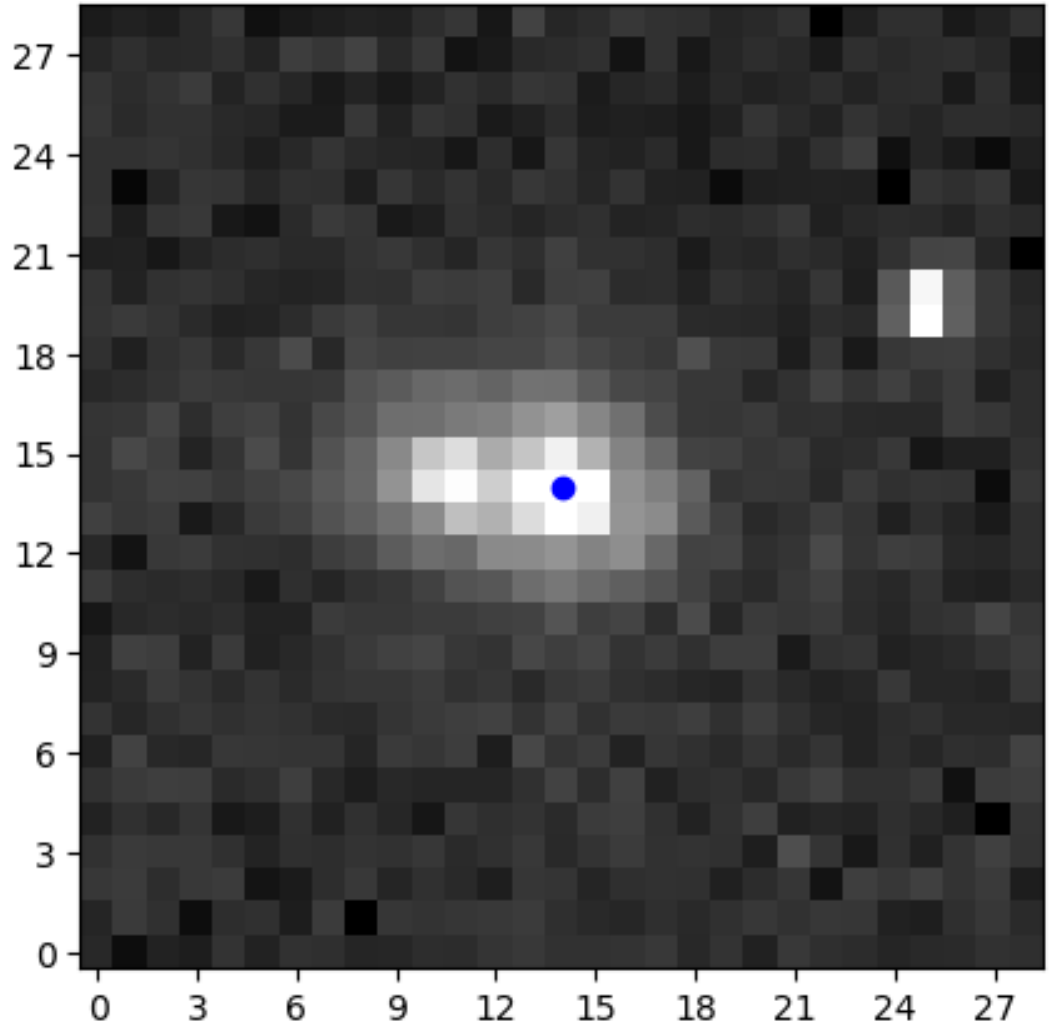}
\includegraphics[width=0.45\textwidth]{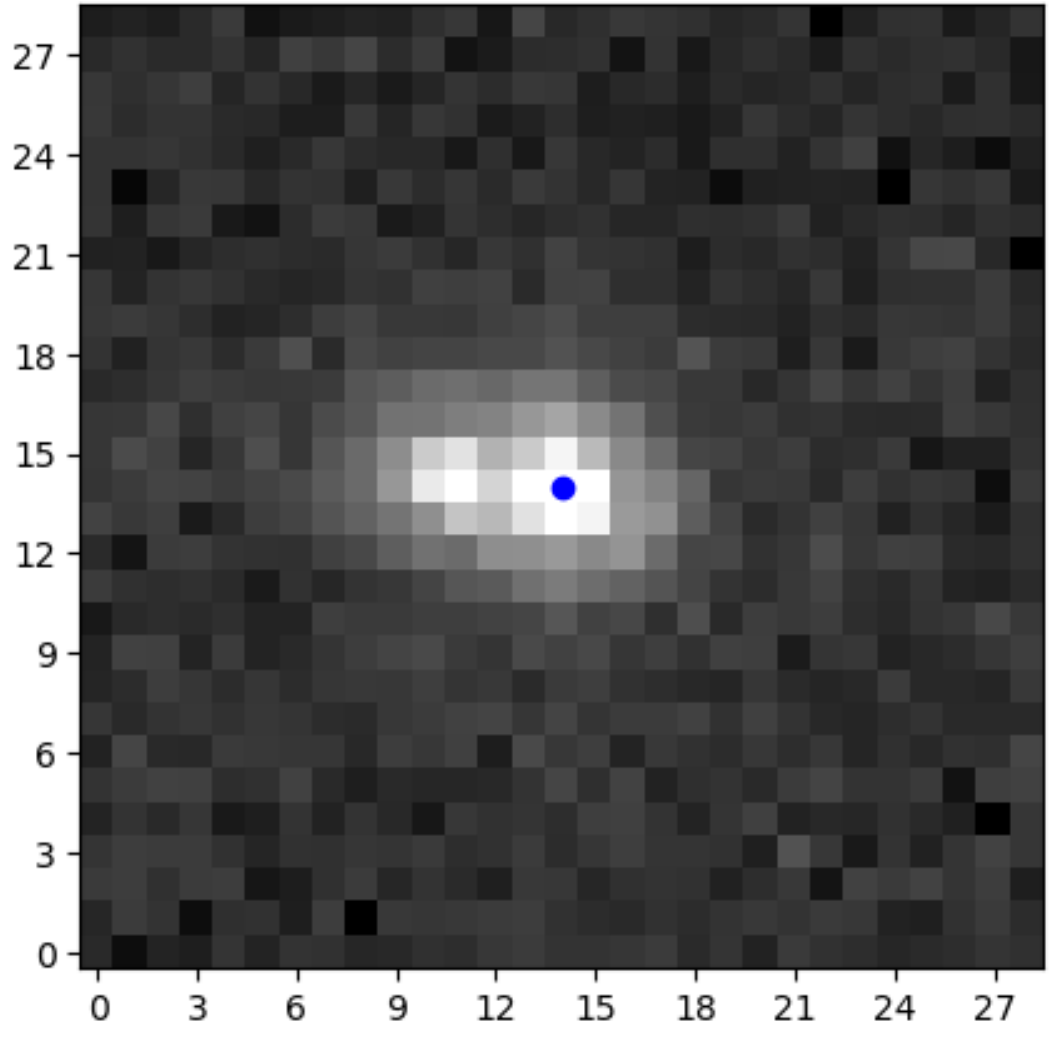}
\caption{\textit{HST}/ACS \textit{type 2} star tile before (left) and after (right) L.A. CR removal, in the filter F580LP. The blue dot mark the central pixel of the tile. The base of each tile is 1.5 arcsec.}
\label{fig:LACR}
\end{figure*}

The presence of cosmic rays (CRs) in the initial tile can heavily affect both the alignment process and later the identification of a  candidate companions in the \textit{residual tile} - see Sec. \ref{Sec:KLIP PSF subtraction and candidate detections}).
Therefore, 
two options are available to mask pixels suspected to be affected by a CR:
\begin{itemize}
\item \textit{DQ mask}: if the occurrence of a CR has  been recorded in the DQ image layer of the \textit{input image}, a median filter around the flagged pixel is applied. For each CR-flagged pixel in the temporary tile, the median value of the counts in a 3 by 3 pixel mask will be evaluated (ignoring any CR-flagged pixel in the mask) and attributed to the pixel to be corrected.
\item \textit{L.A. cosmic ray removal}: cosmic rays can also be identified and removed using the \textit{Astro-SCRAPPY}\footnote{available at \url{https://github.com/astropy/astroscrappy}} python module (\cite{McCully2018} - based on the L.A. Cosmic algorithm from \cite{vanDokkum2001}). Figure \ref{fig:LACR} show of this module performs on a typical tile: the CR is identified and removed without affecting the central resolved binary (labeled \textit{type 2})
\end{itemize}

After CR-cleaning, a second pass is performed to better estimate the centroid of each source. For this step the user can choose between three different strategies:
\begin{itemize}
    \item \textit{maximum}: the brightest pixel within $n$ pixels (default 3 pixels) from the center of the temporary tile is selected as the new center of the tile;
    \item \textit{centroid}: the routine looks for a centroid within $n$ pixels (default 3 pixels) from the center of the temporary tile, assign to its coordinates the new center of the tile. If selected as default, this approach is usually overruled by the \textit{maximum} for any known unresolved double in the catalog (type 2);
    \item \textit{no corrections}: the initial input coordinates are maintained,  no further corrections is applied.
\end{itemize}
By default, the pipeline 
uses the \textit{maximum} option, but any other options can be selected for all entries in the \textit{multiple visits target} dataframe or   (parsing the \textit{average ID} for individual targets when their behaviour is different from the typical set. This maximizes the versatility allowing to adopt the best strategy, given the characteristics of each source and the signal-to-noise ration of each detection.

When the final tile is created, after the centering step, the routine evaluates the new coordinates (if any correction has been made) moving the center of the tile to the new positions in the \textit{input image} and updating the position-labeled columns in the \textit{multiple visits target}  dataframe accordingly. The original sub-pixel position of the source is preserved during this process.
Once the final positions are established, a datacube comprising the following three (four if the option to remove CR has been enabled) different tiles is saved for each target:
\begin{itemize}
\item the tile itself containing science data, labeled as \textit{Data};
\item the corresponding tile cut from the \textit{ERR} layer of the \textit{input image}, labeled \textit{eData};
\item the similar tile obtained from the \textit{DQ} layer of the \textit{input image}, labeled \textit{DQ};
\item if the option to remove CR has been enabled, the CR free data tile is also recorded with label \textit{CR\_Clean}
\end{itemize}

It is advised to carefully inspect each newly created tile before moving to the next step. For this purpose,
routines have been developed to facilitate visual inspection and extract useful information at a glance.
These routines can be easily accessed through a notebook.
A median cube is also saved for each target and filter, containing the median of all the visit tiles and, if the option has been selected, also the cosmic ray-free version of it.


\section{Steps 6 and 7: PSF subtraction and candidate detection}
\label{Sec:KLIP PSF subtraction and candidate detections}
The pipeline performs PSF subtraction using a pipeline derived from pyKLIP\footnote{\url{https://pyklip.readthedocs.io/en/latest/}}  \citep{Wang2015} modified to perform RDI of many sources at once. The pyKLIP module is a python implementation of the KLIP algorithm, invented to achieve accurate PSF subtraction through an operations that can be represented as \cite[for more details see][]{Soummer2012}:
\begin{equation}
    T(n)-\hat{I}_{\psi_0}=\epsilon A(n).
\end{equation}
Here $T(n)$, the target image, is the $n-th$ input tile (either \textit{Data} or \textit{CR\_Clean} label); $\hat{I}_{\psi_0}$ is the best representation of $I_{\psi_0}$, the model of an \textit{isolated} source with no other astronomical signal in the search area $\mathcal{S}$); 
$\epsilon A(n)$ is the extra astronomical source that may ($\epsilon =1$) or may not ($\epsilon =0$) be present in the input tile $T(n)$. 

\begin{figure*}
    \centering
    \includegraphics[width=0.90\textwidth]{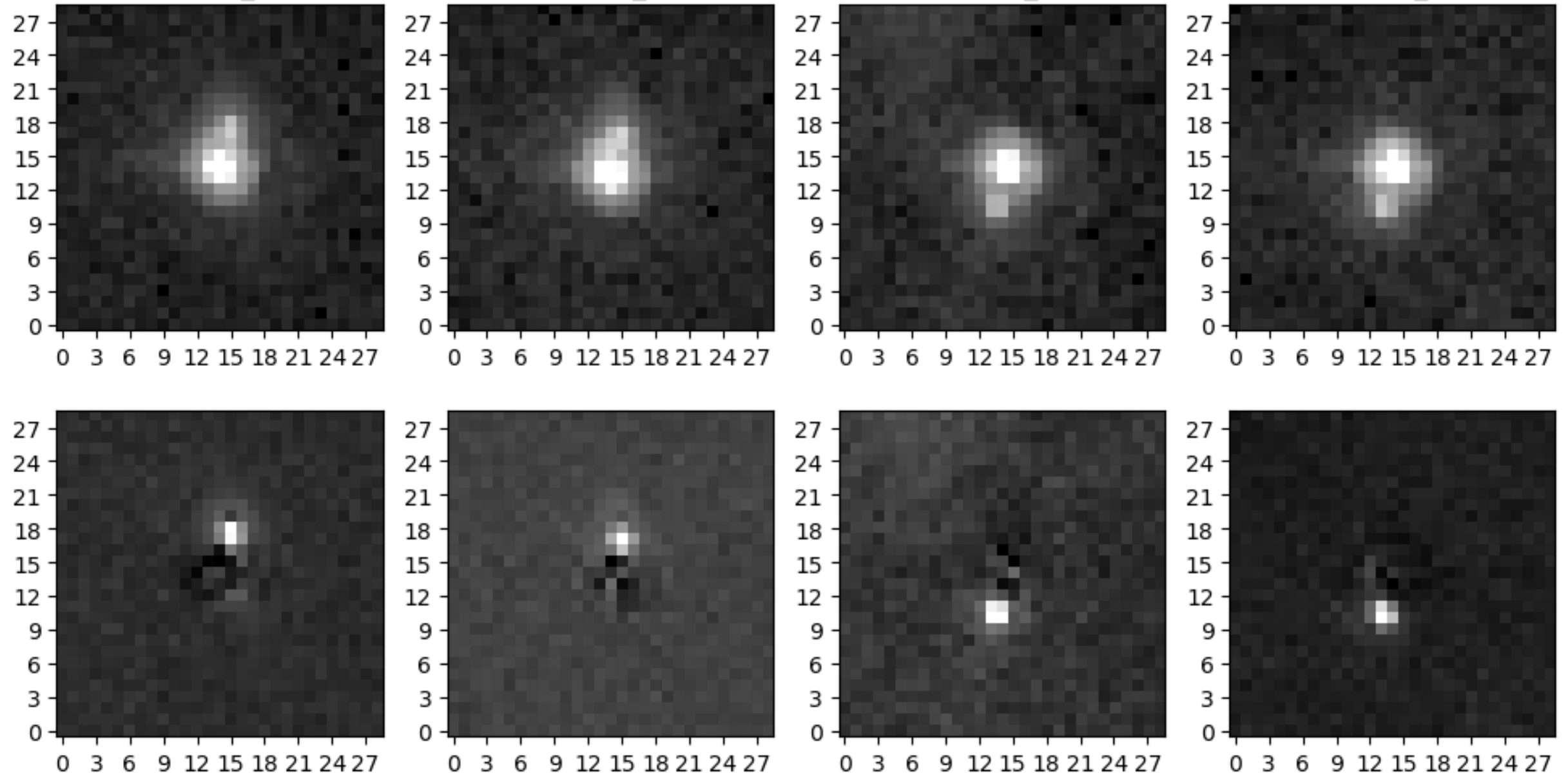}
    \caption{Top row: \textit{HST}/ACS F850LP input tiles of same target observed in 4 different visits with 2 different orientation angle of the telescope. Lower row: PSF subtraction output (\textit{residual's tile}). The additional astronomical signal in the input tile is perfectly clear (once we remove the central star) and it appear as an extended PSF.}
    \label{fig:ACS_sub}
\end{figure*}

\begin{figure*}
    \centering
    \includegraphics[width=0.95\textwidth]{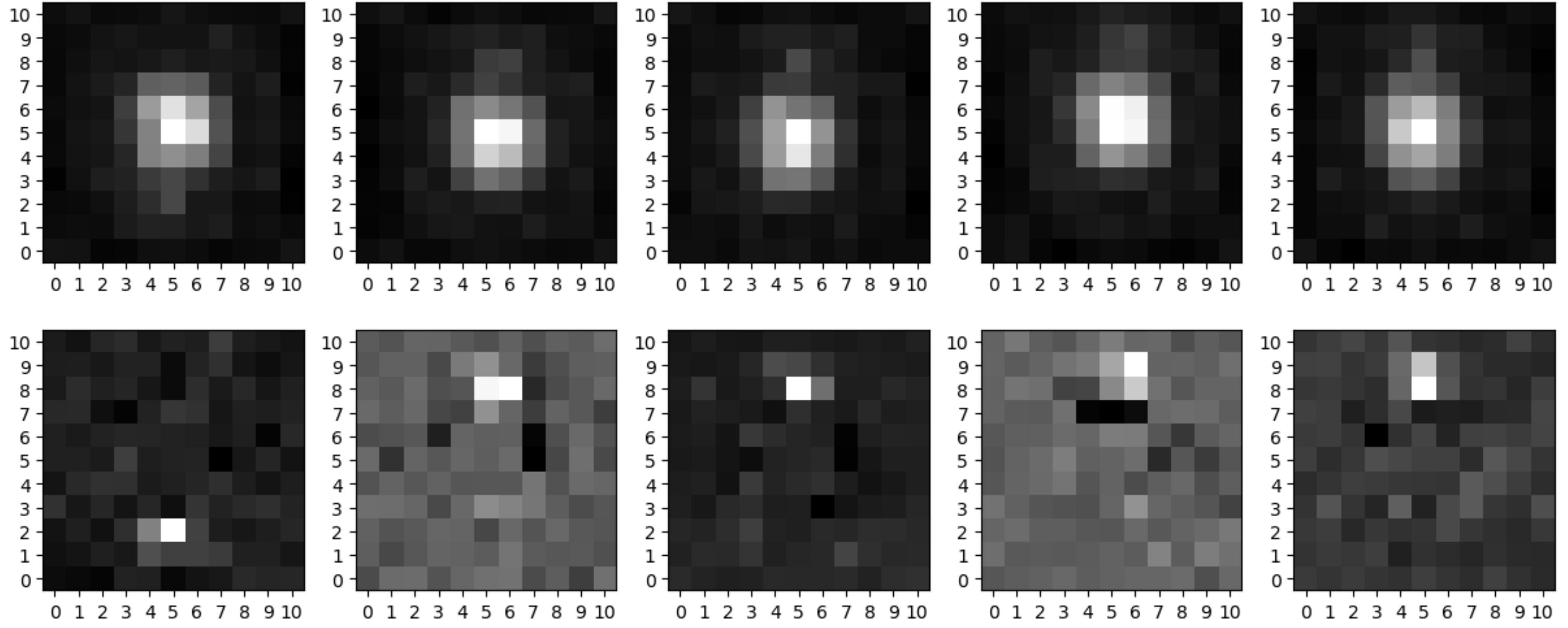}
    \caption{Same as Figure \ref{fig:ACS_sub}. Top row: \textit{HST}/WFC3-IR F130M input tiles. Lower row: PSF subtraction output. In this case the additional astronomical signal in the input tile appears as just a few bright pixels in the \textit{residual's tile}.}
    \label{fig:WFC3-IR_sub}
\end{figure*}

While the ``true'' PSF to be subtracted cannot be known, a set of reference stars can be regarded as random realizations. The Karhunen-Lo{\`e}ve transformation consists in using the spatial correlation between these realization to create an orthonormal basis of eigenimages on which the target star can be projected. 
The typical output of this process is a new tile, having the same size as the input tile, containing the residual of the subtraction process with zero mean. Any additional astronomical signal possibly present in $T(n)$ will be apparent in this \textit{residual's tile}. The residual $\epsilon A(n)$ is nearly orthogonal to the main telescope's PSF and therefore minimally affected by the subtraction process. Depending on its brightness and on the sampling, it may appear as an extended PSF (see Figure \ref{fig:ACS_sub}) or just as a few bright pixels (see Figure \ref{fig:WFC3-IR_sub}).

For each input tile, the PSF subtraction process is repeated iterating over different number of eigenvalues, i.e. the number of modes where to truncate the Karhunen-Lo{\`e}ve transformation, KLIPmode $K_{klip}$. Different KLIPmode values can be used to optimally sample different portions of the search area $\mathcal{S}$, smaller values allowing to better retrieve signal very close to the target source than higher one.  

In essence we are carrying out Reference Differential Imaging (RDI) with the reference PSFs coming from stars in the same mosaicing program and in the same neighborhood of the detector focal plane.

The tiles containing the residual of each KLIP subtraction, as well as the ones containing the models of the isolated star, are added to the multiple visits tiles created before. 


To select a candidate detection, the routine inspects all \textit{KLIPmode} layers relative to the same source (observed over different visits and filters) and compares the position of the brightest pixel in each tile. 
A candidate will be flagged if the following three conditions are realized:
\begin{itemize}
    \item the brightest pixel in the \textit{residual tile} is detected with counts above a user-defined threshold (usually a multiple of the spatial standard deviation, calculated after a initial $3\sigma$ cut to eliminate outliers);
    \item the brightest pixel is detected as such at the same sky position, for all filters and visits;
    \item the brightest pixel is present in at least 2 different KLIPmodes. 
\end{itemize}
To determine spatial coincidence between different visits, the routine factors in the telescope rotation and small misalignment, allowing for a user-defined number of pixels, typically 1, of discrepancy. To choose the KLIPmodes, the routine  compares the \textbf{signal-to-noise}  (SNR) for the KLIPmodes that result in a detection, choosing the one with the higher possible SNR and the lower possible KLIPmode at the same time. Once the KLIPmode representative of this candidate is selected, a median tile is created for the residual and added to the median cube of each target. 
 

\section{\textbf{STEP 8: Photometry}}
\label{Sec:The candidate catalog and photometry}
The pipeline offers three different photometry methods: aperture photometry, matched filter photometry and PSF photometry. While aperture photometry always provides an accurate value if the stars are well isolated and there are no companions, the two other methods can provide a more accurate estimate of the flux of a star if the 
Point Spread Function is well sampled and modeled. This is typically the case for ACS and the visible channel of \textit{WFC3}, but not of \textit{WFC3-IR} due to the coarser pixel scale. We therefore illustrate each method with an application to \textit{ACS} simulations, underlining the strengths and flaws and each strategy in dealing with the data produced by the pipeline.

\subsection{Aperture photometry}
\label{sec:Aperture photometry}
\begin{figure*}
    \centering
    \includegraphics[width=0.49\textwidth]{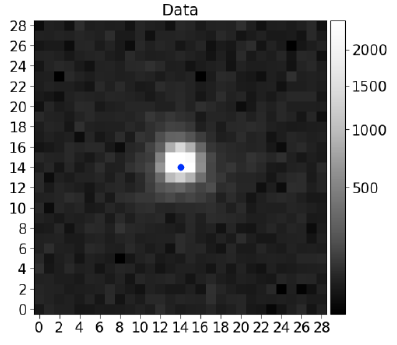}
    \includegraphics[width=0.49\textwidth]{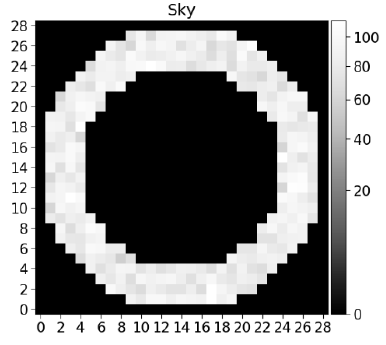}\\
    \includegraphics[width=0.49\textwidth]{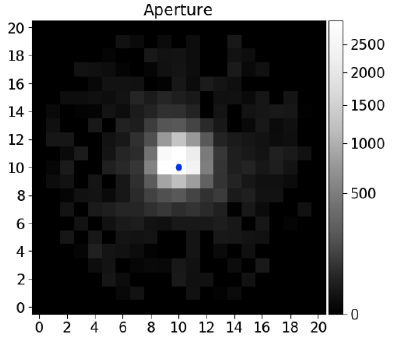}
    \includegraphics[width=0.49\textwidth]{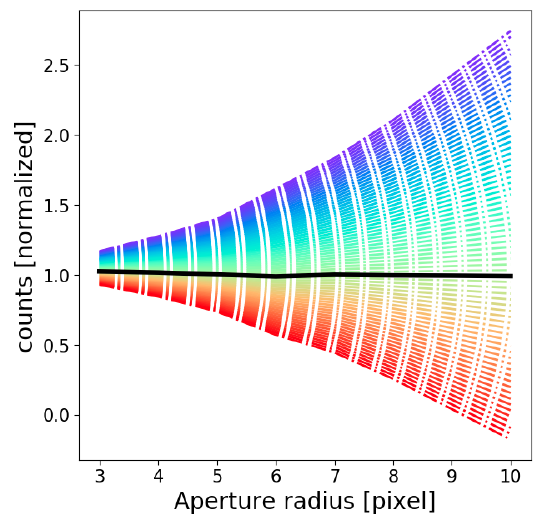}
    \caption{Top row: input data tile (left) and sky annulus  (right). The blue point marks the center of the tile. Bottom row: Sky-subtracted aperture-selected data tile (left) and grow curves (right). Each different curve in this last plot is calculated correcting the estimation of the sky by a step of $\pm 1\%$ to a maximum of $\pm 30\%$ (blue/red). The black line shows the "flattest" curve across all the radii and corresponds to the curve with a sky correction of $-1\%$. Each curve in this example has been normalized to the average counts of the black curve for radii bigger than 5.}
    \label{fig:AP_phot}
\end{figure*}

To perform aperture photometry the pipeline uses the \textbf{\texttt{photometry$_{AP}$}} routine which integrates some of the \textbf{\texttt{photutils}}\footnote{\textbf{\texttt{photutils}} is a python module affiliated to Astropy that provides tools for detecting and performing photometry of astronomical sources. \url{https://photutils.readthedocs.io/en/stable/index.html}} feature to perform aperture photometry.
The two frames at the top of Figure~\ref{fig:AP_phot} show the two tiles required by this routine.
The routine uses the \textbf{\texttt{CircularAnnulus}} module 
of \textbf{\texttt{photutils}} to evaluate the median sky background ($C_{sk}$). Specifically, it takes a $3 \: \sigma$ cut median of the sky in an annulus  between two radii ($r_a$ and $r_b$)  (\textit{Sky} tile in Figure \ref{fig:AP_phot}). This sky estimate is then subtracted from the data tile creating a \textit{sky-subtracted} tile. Then, using the \textit{CircularAperture} module of \textbf{\texttt{photutils}}, the routine sets to zero all pixels outside the circular area of radius $r_{i}$  
creating a final \textit{sky-subtracted aperture} tile (or simply \textit{Aperture} tile in Figure \ref{fig:AP_phot}), whose total counts ($C_{ap}'$) provide the source counts inside $r_{i}$ . 
To check the accuracy of the sky estimate, many \textit{curves of growth} are created (Figure \ref{fig:AP_phot}, bottom-right) correcting the sky value by small amounts and performing multi-aperture photometry at different radii to assess the systematic errors on the sky estimate. 
Indeed, when the sky is correctly estimated then the curve will remain flat over the full range of selected radii. Otherwise, the multi-aperture photomery will show  a parabolic shape due to the fact that the sky error grows quadratically with the number of pixels in the aperture, i.e. $\propto r_i^2$. If the sky has been overestimated, too much flux will be removed as the radius increases and the counts will show a parabolic decrease vs. radius; conversely, if the sky has been underestimated. So, by making different grow curves, the routine looks for the flattest curve in the sample (black curve in Figure 5) and returns the correct sky estimate.

To evaluate the photometric uncertainty we use the classic relation by \cite{Stetson1987}:
\begin{equation}
    \label{eq:ap_error}    
    \Delta C_{ap}=\sqrt{var_1+var_2+var_3}
\end{equation}
where \textbf{\textit{var$_1$},\textit{var$_2$} and \textit{var$_3$} }are the three main source of errors added in quadrature. The first term,
\begin{equation}
    \label{eq:var1}
    var_1=N_{ap} \: std(C_{sk})^2 ,
\end{equation}
represents the random noise floor in the aperture, including readout noise and contamination from neighbouring stars. This term is given by the variance of the sky values, $C_{sk}$, multiplied by the number of pixels in the aperture, $N_{ap}$. The second term accounts for the photon noise associated to the brightness of the source, 
\begin{equation}
    \label{eq:var2}
    var_2=C_{ap}
\end{equation}
where $C_{ap}$ represent the photoelectrons counted inside the aperture, and the third term
\begin{equation}
    \label{eq:var3}
    var_3=N_{ap}^2\:\Bigg(\frac{std(C_{sk})^2}{N_{sky}}\Bigg) ,
\end{equation}
is the uncertainty in the estimate of the sky brightness, $C_{sk}$, not accounted for by the  variance of source counts but still affecting the actual measure. 

Once the final counts in the aperture with their uncertainties have been evaluated, they are  converted in magnitude, $m$, with its relative uncertainty, $dm$, through the relations:
\begin{equation}
    m=-2.5 \: log_{10}\Bigg(\frac{C_{ap}}{ExpTime}\Bigg)+ZP
    \label{eq:mag}
\end{equation}
\begin{equation}
    dm=1.0857\:\Bigg(\frac{\Delta C_{ap}}{C_{ap}} \Bigg)
    \label{eq:emag}
\end{equation}
where \textit{ExpTime} is the exposure time and \textit{ZP} is the zero point.

\subsubsection{Undersampled PSF}

As already noted in \cite{Strampelli2020}, due to the fact that \textit{WFC3-IR} PSF is highly undersampled, most of the flux from a faint candidate companion is contained within just a few pixels of the residual tile. Also, the number of pixel at disposal is so small that it is not possible to assess very precisely the star location and the centering of the aperture is usually rather poor.
Thus, to derive its total flux, one has to apply a large and rather uncertain aperture correction. 
To address this problem, the pipeline performs photometry of the companion using a 4-pixel aperture and  uses the processed data to establish the relative aperture correction. To this purpose, for each isolated source in the input catalog the routine determines the brightest $2 \times 2$ pixel area including the brightest pixel of the original image. After probing the 4 possible area positions, the routine selects the one providing the highest total counts, C$_{4p}$. The known magnitude of the source is then related to the four pixel counts through the equation:
\begin{equation}
\label{eq:m4p}
m_{4p} = -2.5 log_{10}(C_{4p}) + \Delta
\end{equation}
where $\Delta$ is a zero point term relating the 4-pixel sum to the known source photometry.
Repeating this calculation over multiple ``standard'' stars, the final $\Delta$ value is determined as the $3\sigma$ cut median of the full set, with an uncertainty $sigma_\Delta$ given by the associated standard deviation.

Once the final $\Delta$ have been determined, a similar 4-pixel aperture is applied to the residual tile. 
Equations \ref{eq:m4p}, with the newly evaluated $C_{4p}$ counts for the companion and the final $\Delta$ value, 
thus provide the magnitudes of the candidate companions. The associated uncertainties  are similar to those evaluated for a generic aperture photometry (see paragraph \ref{sec:Aperture photometry}) with the added uncertainty on the aperture correction.
Note that estimating the 4-pixel aperture photometry of a companion, the \textit{sky} refers to the measured background in the residual tile.

\subsection{Matched filter photometry}
\label{sec:Matched filter photometry}
For well sampled data, the pipeline offers a second method to extract photometry, \textit{matched filter photometry (MF)}  \citep{Rothstein1954,Turin1960}. This approach solves the problem of detecting with the highest \textbf{signal-to-noise} a signal of known shape in noisy data. The solution is given by the cross-correlation between a known signal template (the reference PSF in this case) and an unknown noisy signal (the target). The pipeline implements this method through the \textbf{\texttt{photometry$_{AP}$}} routine which takes advantage of the MF routines present in \textbf{\texttt{pyKLIP}} package \citep{Ruffio2017}. These perform the convolution of the template with an image using the scipy signals processing library of Fast Fourier Transformations. 
\begin{figure*}[!t]
    \centering
    \includegraphics[width=0.99\textwidth]{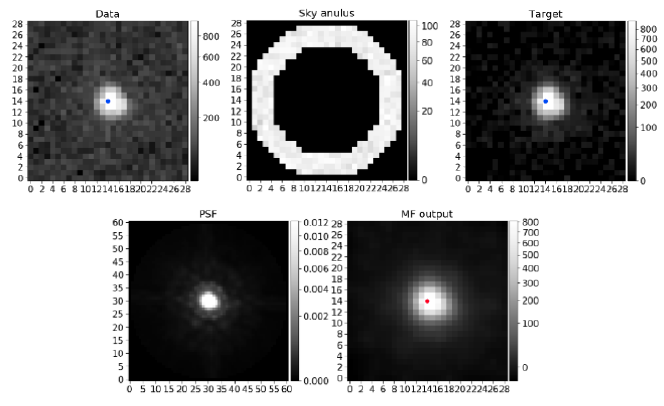}\\
    \caption{Top row: input data tile (left), sky annulus (center) and sky subtracted target tile (right). Bottom row: PSF reference tile from Tiny Tim (left) and MF output tile (right). The red (blue) dot marks the maximum (center) in tile.}
    \label{fig:MF_phot}
\end{figure*}

Figure \ref{fig:MF_phot} shows the different tiles used by the \textbf{\texttt{photometry$_{AP}$}} routine. 
First, the sky is evaluated using the same approach adopted by the \textbf{\texttt{photometry$_{AP}$}} routine, adopting an annulus between r$_a$ and r$_b$ (\textit{Sky annulus} tile in Figure \ref{fig:MF_phot}). Also in this case, the sky estimate can be further refined using the \textit{grow curves} as explained in paragraph \ref{sec:Aperture photometry}. The sky is then subtracted from the \textit{Data} tile to produce a \textit{sky-subtracted} target tile (or just \textit{Target} in Figure \ref{fig:MF_phot}) 
This tile is convolved with the reference tile, i.e. the \textit{PSF} tile obtained for example from Tiny Tim\footnote{Tiny Tim is a program that generates simulated Hubble Space Telescope point spread functions (PSFs). \url{https://www.stsci.edu/software/tinytim/}} and the output of this operation is stored in the \textit{MF output} tile. 

The counts of the star can then be recovered dividing the brightest pixel in the MF output 
by the \textit{throughput} (i.e. the normalization factor that will rescale the counts MF output to the real flux of the star):

\begin{equation}
    \label{eq:MF_counts}
    C_{MF}=\frac{max(MF_{output})}{throughput}
\end{equation}
\begin{equation}
    \label{eq:MF_thpt}
    throughput=|\Sigma_{i,j} \: abs(PSF_{i,j})^2|
\end{equation}
where $PSF$ represents the input template tile.
To estimate the uncertainties on C$_{MF}$, the routine follows the same equations \ref{eq:var1} to \ref{eq:var3} used for \textbf{\texttt{photometry$_{AP}$}} with the difference that since no aperture is performed in this photometry, a "noise-equivalent area" needs to be defined from to derive the  number of pixels ($\overline{N}_{ap}$) contributing to the uncertainty estimate.

\begin{figure*}[!t]
    \centering
    \includegraphics[width=0.32\textwidth]{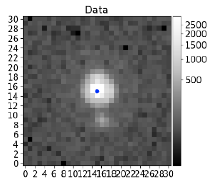}
    \includegraphics[width=0.32\textwidth]{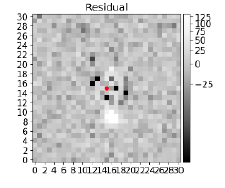}
    \includegraphics[width=0.32\textwidth]{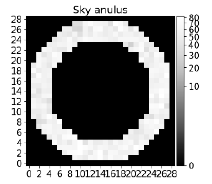}\\
    \includegraphics[width=0.32\textwidth]{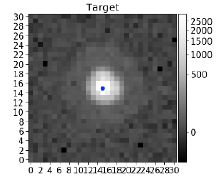}
    \includegraphics[width=0.32\textwidth]{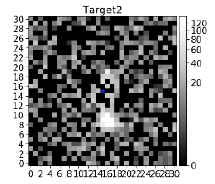}\\
    \includegraphics[width=0.64\textwidth]{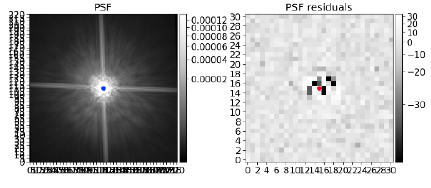}
    \includegraphics[width=0.30\textwidth]{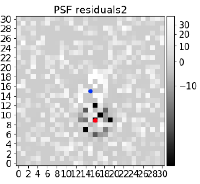}\\
    
    \caption{
    Top row: input data tile (left), residual tile after PSF subtraction (central) and  sky annulus (right). Central row: isolated sky-subtracted primary (left) and isolated residual-background-subtracted companion (right). Bottom row: PSF reference tile from Tiny Tim and PSF residual tile for primary and companion. The blue (red) dot marks the coordinate for PSF fitting.}
    \label{fig:PSF_phot}
\end{figure*}

\begin{figure*}[!t]
    \centering
    \includegraphics[width=0.64\textwidth]{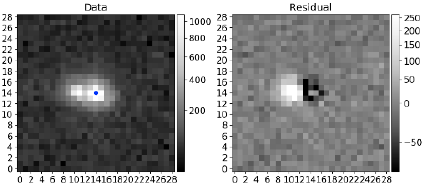}
    \includegraphics[width=0.32\textwidth]{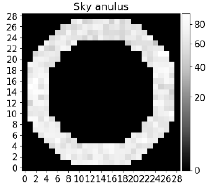}\\
    \includegraphics[width=0.32\textwidth]{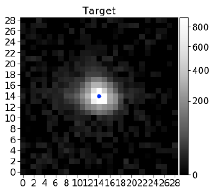}
    \includegraphics[width=0.32\textwidth]{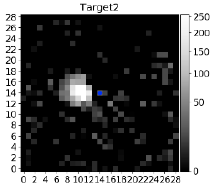}\\
    \includegraphics[width=0.64\textwidth]{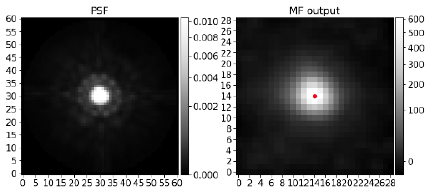}
    \includegraphics[width=0.32\textwidth]{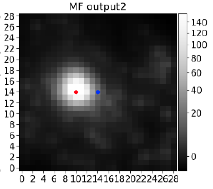}
    \caption{Top row: input data tile (left), residual tile after PSF subtraction (center) and sky annulus (right). Central row: isolated sky-subtracted primary (left) and isolated residual-background-subtracted companion (right). Bottom row: PSF reference tile from TinyTim and MF output tile for primary and companion. The blue (red) dot marks the center (maximum) in the MF output tile.}
\label{fig:MF_phot_comp}
\end{figure*}

Following \cite{King1983}, the effect of the background, $B$, can be computed by adding it over an equivalent area whose dimension depends on the size and shape of the PSF. The pipeline uses King's approximate rule of thumb for the equivalent noise area 
\begin{equation}
    \label{eq:eq-area}
    \sigma^2_{B}=8\pi a^2B
\end{equation}
where $a\sim 0.5\lambda/D$ is half the diffraction-limited angular resolution of the telescope at the effective wavelength of the filter in use. Future versions will include the more rigorous results presented by \cite{King1983} in tabular form. Once $\sigma^2_B$ is evaluated, it can be combined witht he Poisson noise associated to the brigthness of the star as:
\begin{equation}
    \label{eq:MF error counts}
    \Delta C_{MF}=\sqrt{C_{MF}+\sigma^2_B}
\end{equation}

Once C$_{MF}$  and  $\Delta C_{MF}$ are evaluated, the same Eq. \ref{eq:mag} and \ref{eq:emag} can be applied.

\subsection{Matched Filter photometry of binaries}
In the case of a close binary system, where aperture photometry fails due to the presence of both stars in the extraction aperture, the MF photometry technique can be easily applied to retrieve the flux of primary and companion apart.
With reference to the tiles shown in Figure \ref{fig:MF_phot_comp}, to retrieve the flux of the primary, the routine subtracts the \textit{residual tile} from the \textit{input tile} (Data) obtaining a new tile (target) where the primary is isolated and the \textbf{\texttt{photometry$_{AP}$}} routine can work as explained in the previous paragraph.
For the companion, instead, the routine evaluates and subtracts the  $3\sigma$ cut median background in the \textit{residual} tile producing a new tile (target2) for the \textbf{\texttt{photometry$_{AP}$}}.


\subsection{PSF photometry} 
The third option to perform photometry of well sampled data is PSF photometry. The  \textbf{\texttt{photometry$_{PSF}$}} routine performs PSF fitting making extensive use of the \textbf{\texttt{BasicPSFPhotometry}} package from \textbf{\textbf{\texttt{photutils}}}\footnote{\url{https://photutils.readthedocs.io/en/stable/index.html}}. As the initial guess for the coordinates and flux, \textbf{\texttt{photometry$_{PSF}$}} evaluates a centroid on the tile and uses the flux estimated by \textbf{\texttt{photometry$_{AP}$}}.
Once the flux of the star has been evaluated through PSF fitting, \textbf{\texttt{photometry$_{PSF}$}} takes advantage of equations similar to Eq. \ref{eq:eq-area} and Eq. \ref{eq:MF error counts} to estimate the uncertainties on the counts. Then it convert counts and errors to magnitude units using Eq. \ref{eq:mag} and \ref{eq:emag}. Moreover, a reduced $\chi^2$ is evaluated for each performed photometry as:
\begin{equation}
    \begin{split}
        \chi^2=\frac{1}{m-n}\sum_{i,j}\Bigg(\frac{x_{i,j}-\mu_{i,j}}{\sigma_{i,j}}\Bigg)^2=\\=\frac{1}{m-n}\sum_{i,j}\Bigg(\frac{Residual_{i,j}}{eData_{i,j}}\Bigg)^2
    \end{split}
\end{equation}
where \textit{m} is the number of number of pixels in the tile, \textit{n} is the number of free parameter for the fit, typically $n=3$ fitting both the positions \textit{i,j} and the flux, $x_{i,j}-\mu_{i,j}$ is the difference between the observable ($x_{i,j}$) and the model ($\mu_{i,j}$). A \textit{residual tile} is produced by the \textbf{\texttt{photometry$_{PSF}$}}, together with  
a tile comprising the pixel-related uncertainties on the flux of the observable, corresponding to the \textit{eData tile} created during the step 4 of the pipeline, see Sec.~\ref{sec:Making the tiles}
and Figure \ref{fig:PSF_phot} for a graphical example.


\subsection{PSF photometry of binaries}
Like in the case of \textit{matched filter photometry}, the \textbf{\texttt{photometry$_{PSF}$}} routine can be utilized to estimate the flux of both members of a binary system once it has been split in its two components through KLIP. The routine will perform steps similar to \textbf{\texttt{photometry$_{AP}$}}, as shown in Figure \ref{fig:PSF_phot} , evaluating the flux for both primary and companion through PSF fitting instead of performing a convolution. 
As explained before, the \textit{PSF residual} tiles can be utilized to evaluate the goodness of the performed fit for both the primary and companion.

\subsection{Pipeline photometry performance}
\label{sec:Pipeline photometry performance}
Even though the \textbf{\texttt{photometry$_{AP}$}} package is included in the pipeline, it's not fully integrated yet, so we advise the user to use it with caution until a new version of the pipeline is released.
In the following, we will limit our tests only to the \textbf{\texttt{photometry$_{AP}$}} and \textbf{\texttt{photometry$_{PSF}$}} routines.

The performance of the latest two photometry suites has been tested for the \textit{HST}/ACS instrument simulating isolated stars and binaries with different separation and magnitudes difference. Given our primary scientific interest on the Orion Nebula Cluster (ONC), we focused on systems that could be plausibly observed in the ONC. 
The results obtained with \textit{aperture}, \textit{matched filter}  and \textit{PSF photometry} were analyzed to determine photometric accuracy and the possible presence of systematic errors. 

Similar test have been conducted also for the branch of the pipeline working with \textit{HST}/WFC3-IR data. Since they have been extensively presented in \citep{Strampelli2020}, in the following paragraphs  we will mainly focus on \textit{HST}/ACS data simulations. Given the better spatial sampling of ACS vs. WFC-IR, our results may provide a more accurate assessment of the advantages and limitations of each method.

\subsubsection{Performance on isolated stars}
\label{sec:Isolated stars}
To test the performance of the \textbf{\texttt{photometry$_{AP}$}} and \textbf{\texttt{photometry$_{PSF}$}} routines, thousands of isolated stars have been generated to closely resemble the stars present in the Orion Nebula Cluster both in flux and position on the detector. 

Using Tiny Tim, we created 50 PSF datacubes in five different target filters: F435W, F555W, F658N, F775W and F850LP, 
with the sources equally spaced on both chips of the ACS/WFC focal plane
Each PSF datacube was created assuming effective temperatures between 2000 and 3000 $^\circ$K in steps of 250 $^\circ$K 
, appropriate to sample the lower end of the temperature range of young low-mass companions and M-dwarfs. 

Each PSF datacube comprises 25 different sub-pixel shifted PSFs created with the standard Tiny Tim \textit{SUB=5} option, which splits each native pixel in 25 sub-pixels. When the \textit{SUB} option is enabled, Tiny Tim does not convolve the PSF with the charge diffusion (CD) kernel. Following \cite{Hoffmann2017} we applied the CD kernel to the oversampled PSF, before shifting the PSF  by a finite amount of sub pixels, and rescaled to the native resolution.
This operation consist in applying the CD kernel to subsets of the PSF tile that posses the same pixel phase. Arranging the subsampled pixel in this way places neighboring pixels in an order similar to the order of the natural ACS resolution. Then, each sample of subsampled pixel with the same pixel phase is convolved with the CD kernel. This process was repeated 25 times to cover all the possible pixel phases in the 5x5 subsampled PSF. 

Having convolved by the CD kernel the subsampled PSF, a subpixel shift has been applied to center the PSF in each of the 25 sub-pixels and the PSF is resampled to the native ACS resolution producing 25 slightly different shifted PSF for each temperature, chip and position on the CCD. 
\begin{table*}
	\begin{center}
		\begin{tabular}{c|c|c|c|c|c}
			  & F435W & F555W & F658N & F775W & F850LP \\
			\hline
			b$_{ap}$ & 0.003 $\pm$ 0.061 & -0.006 $\pm$ 0.143 & -0.007 $\pm$ 0.262 & -0.010 $\pm$ 0.106 & -0.004 $\pm$ 0.099 \\
			b$_{psf}$ & 0.001 $\pm$ 0.017 & 0.001 $\pm$ 0.019 & -0.000 $\pm$ 0.246 & -0.000 $\pm$ 0.102 & 0.000 $\pm$ 0.095 \\
		\end{tabular}
	\end{center}
    \caption{Values of the constants evaluated from each simulation as explained in the text, with correspondent standard deviation of all points.}
    \label{tab:fit_s}
\end{table*}
\begin{figure*}[!t]
    \centering 
    \includegraphics[width=0.31\textwidth]{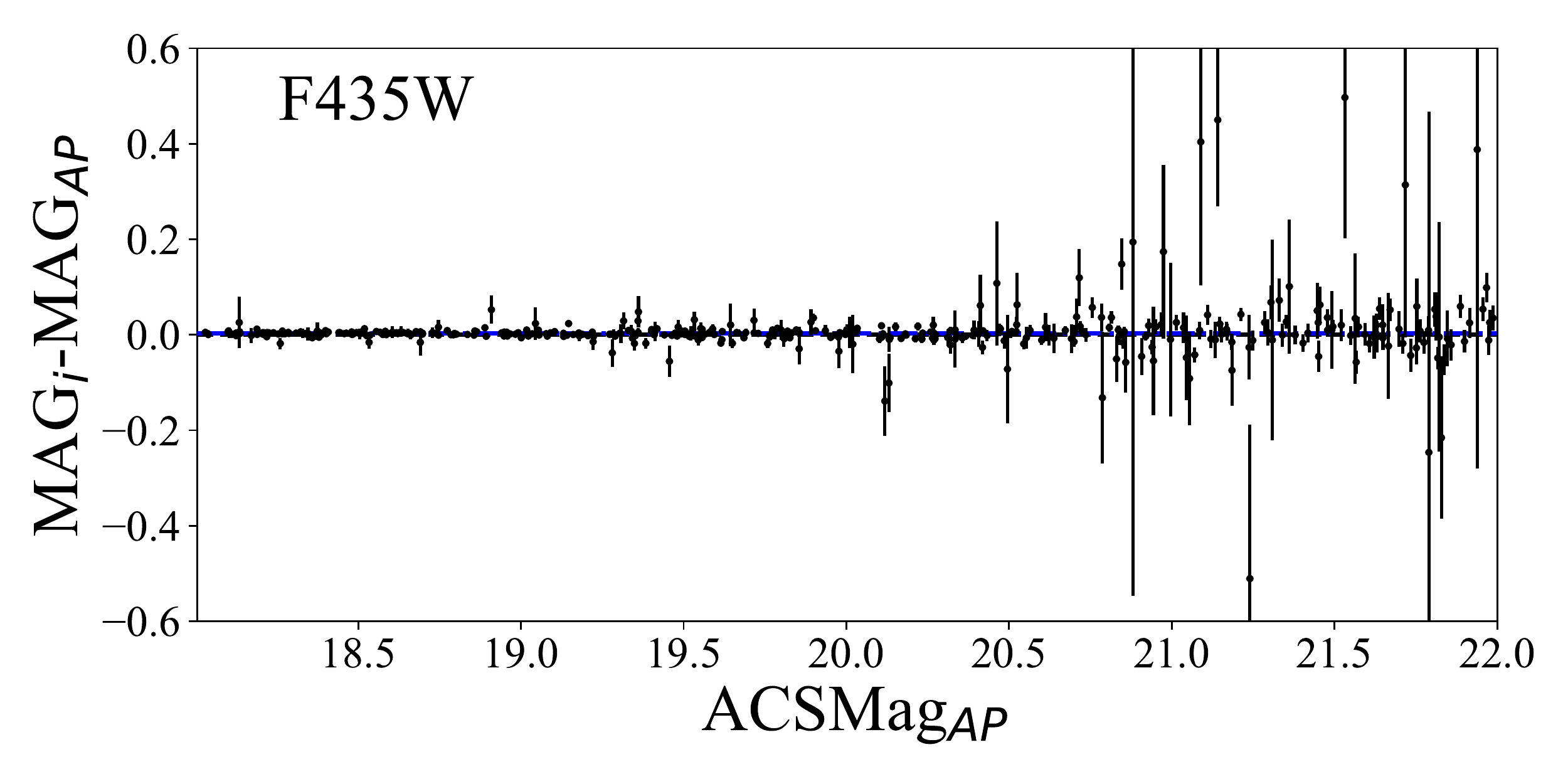}
    \includegraphics[width=0.31\textwidth]{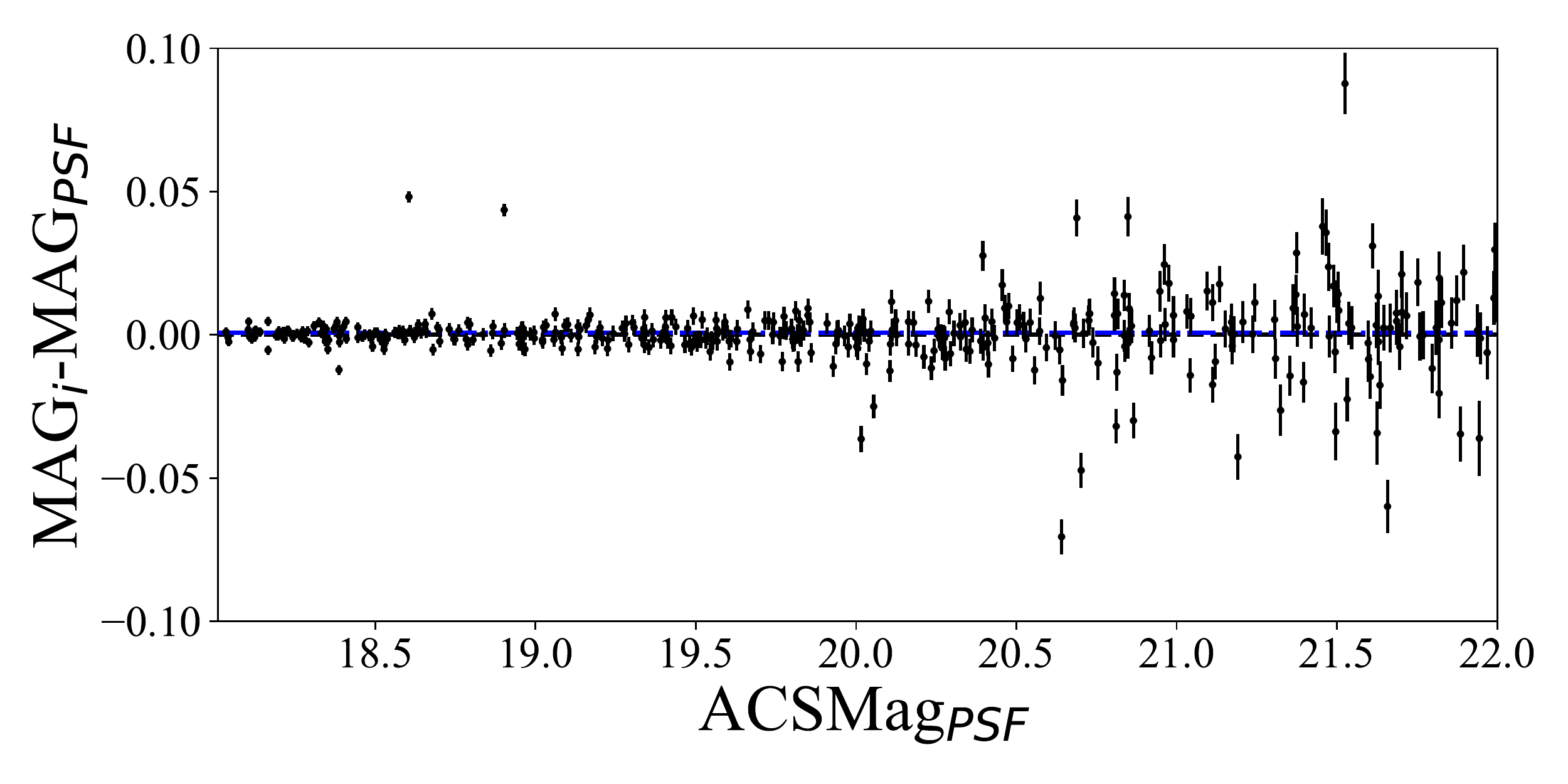}\\
    \includegraphics[width=0.31\textwidth]{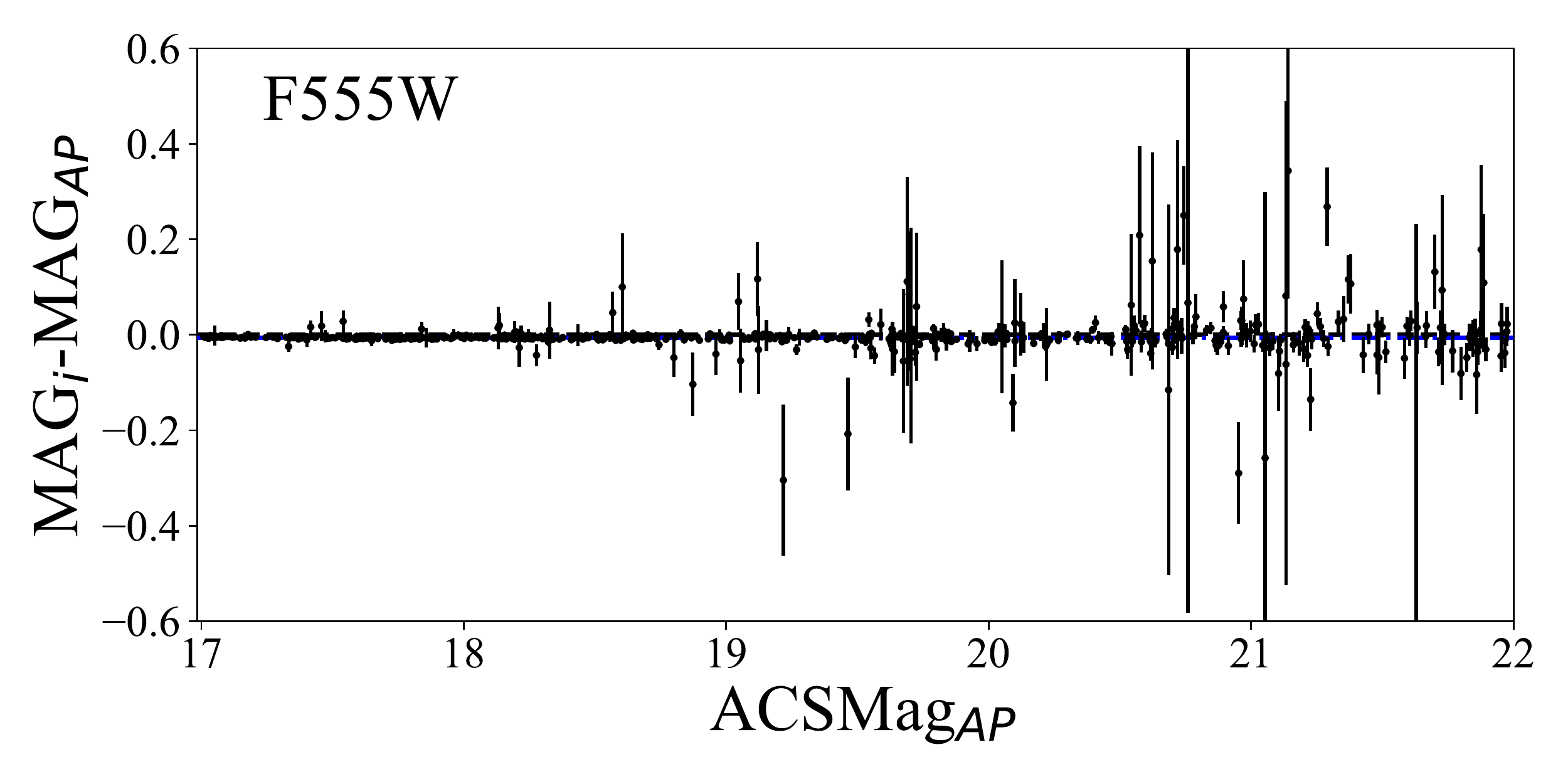}
    \includegraphics[width=0.31\textwidth]{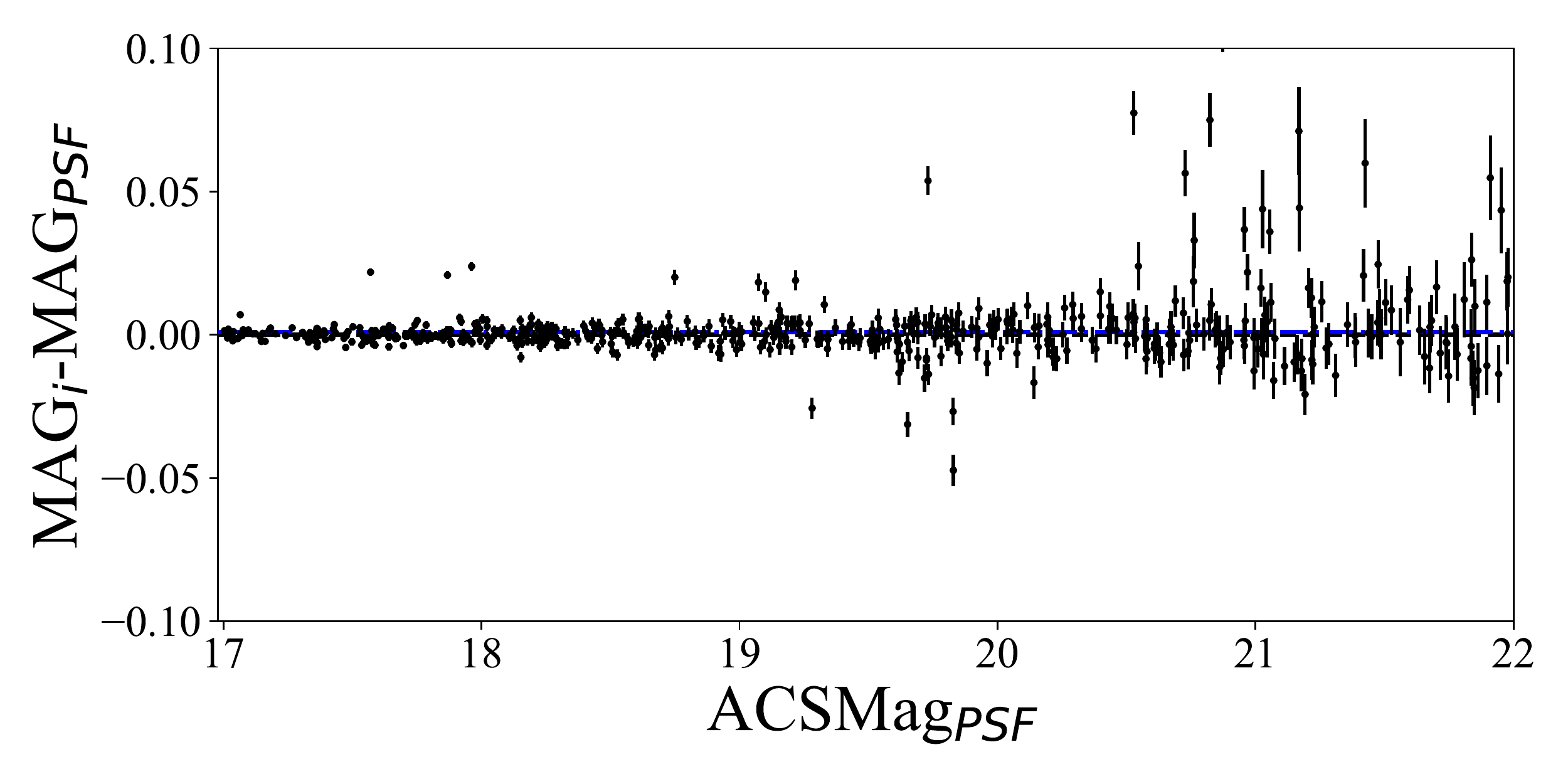}\\
    \includegraphics[width=0.31\textwidth]{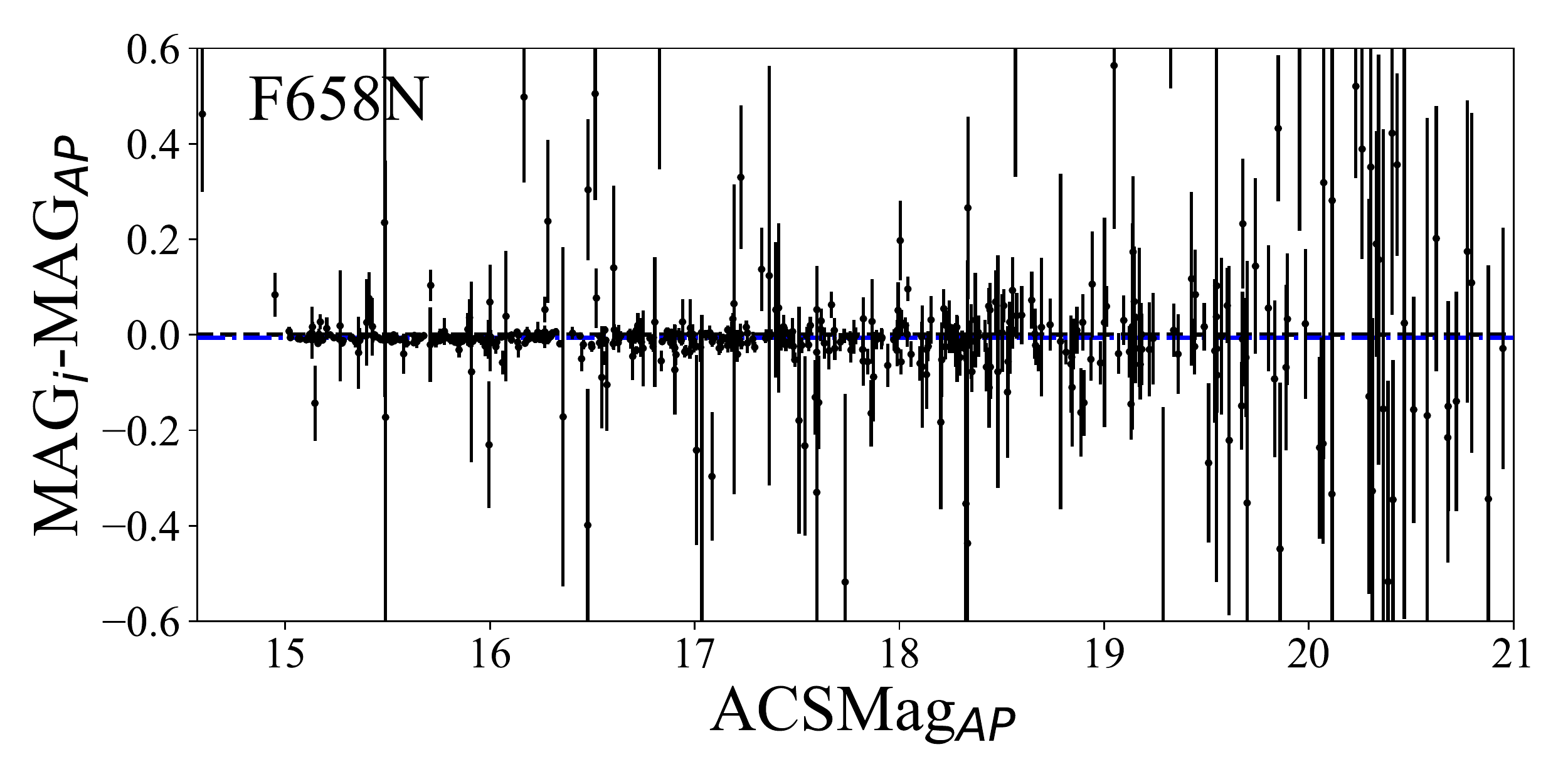}
    \includegraphics[width=0.31\textwidth]{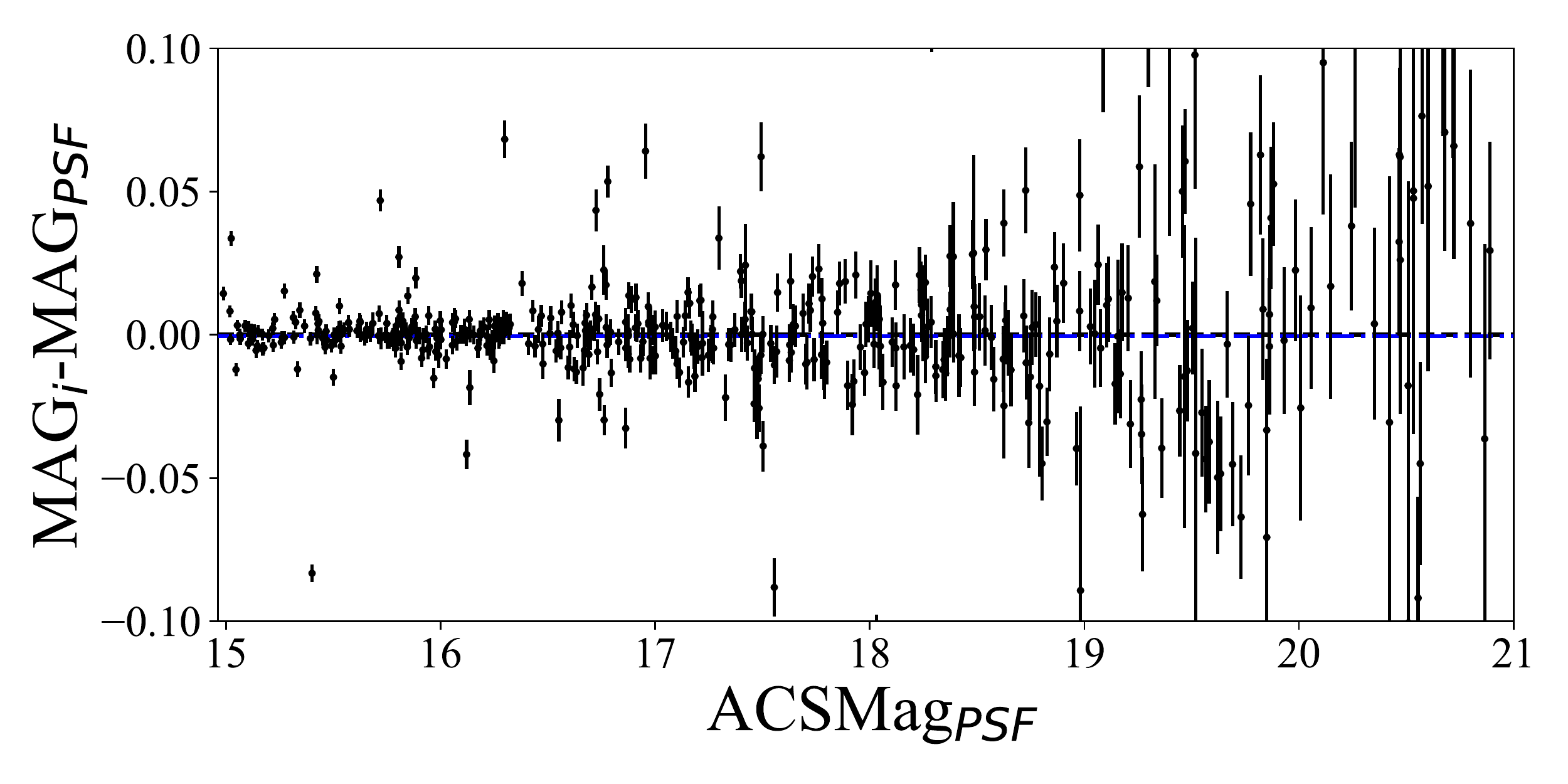}\\
    \includegraphics[width=0.31\textwidth]{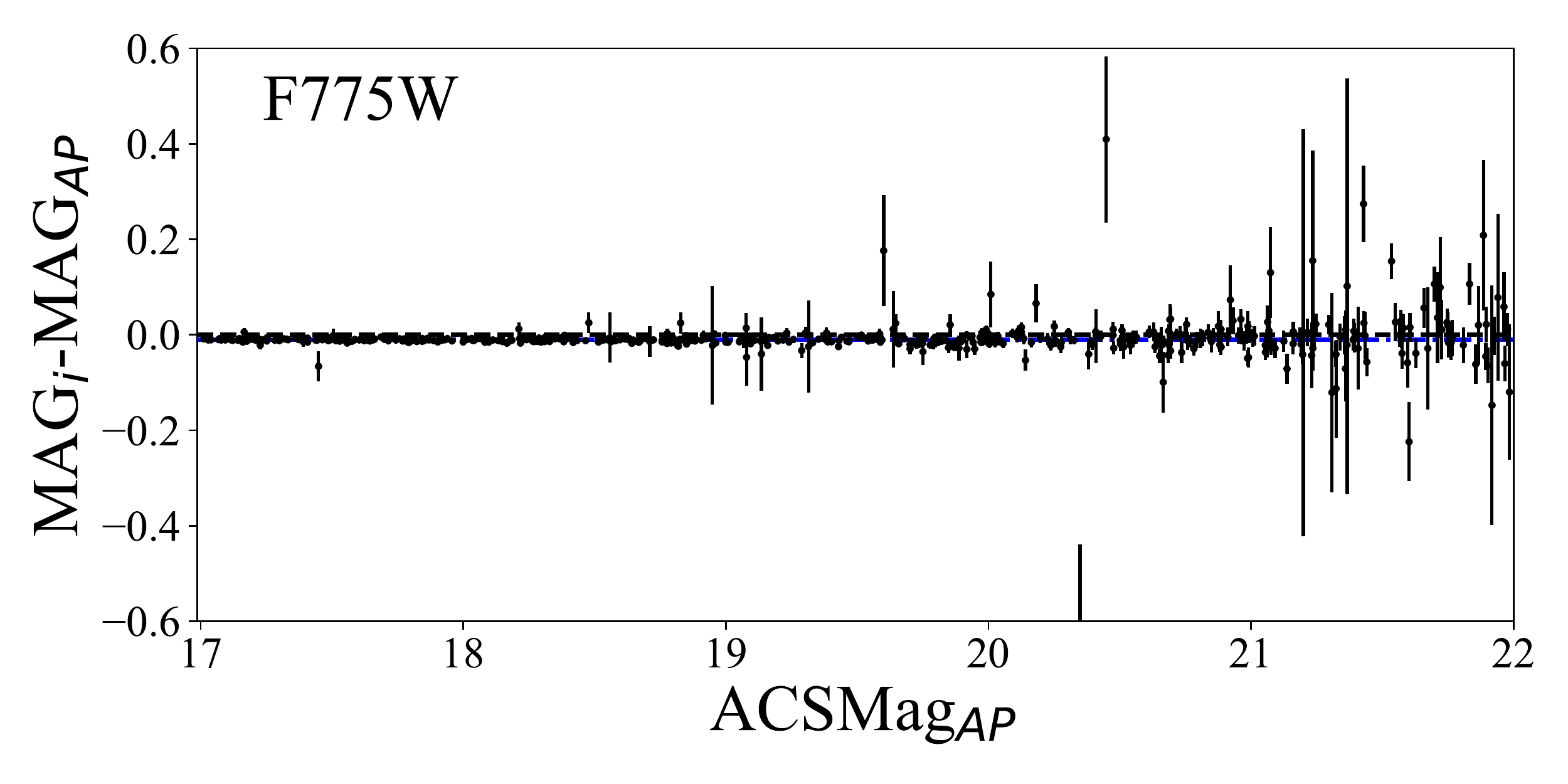}
    \includegraphics[width=0.31\textwidth]{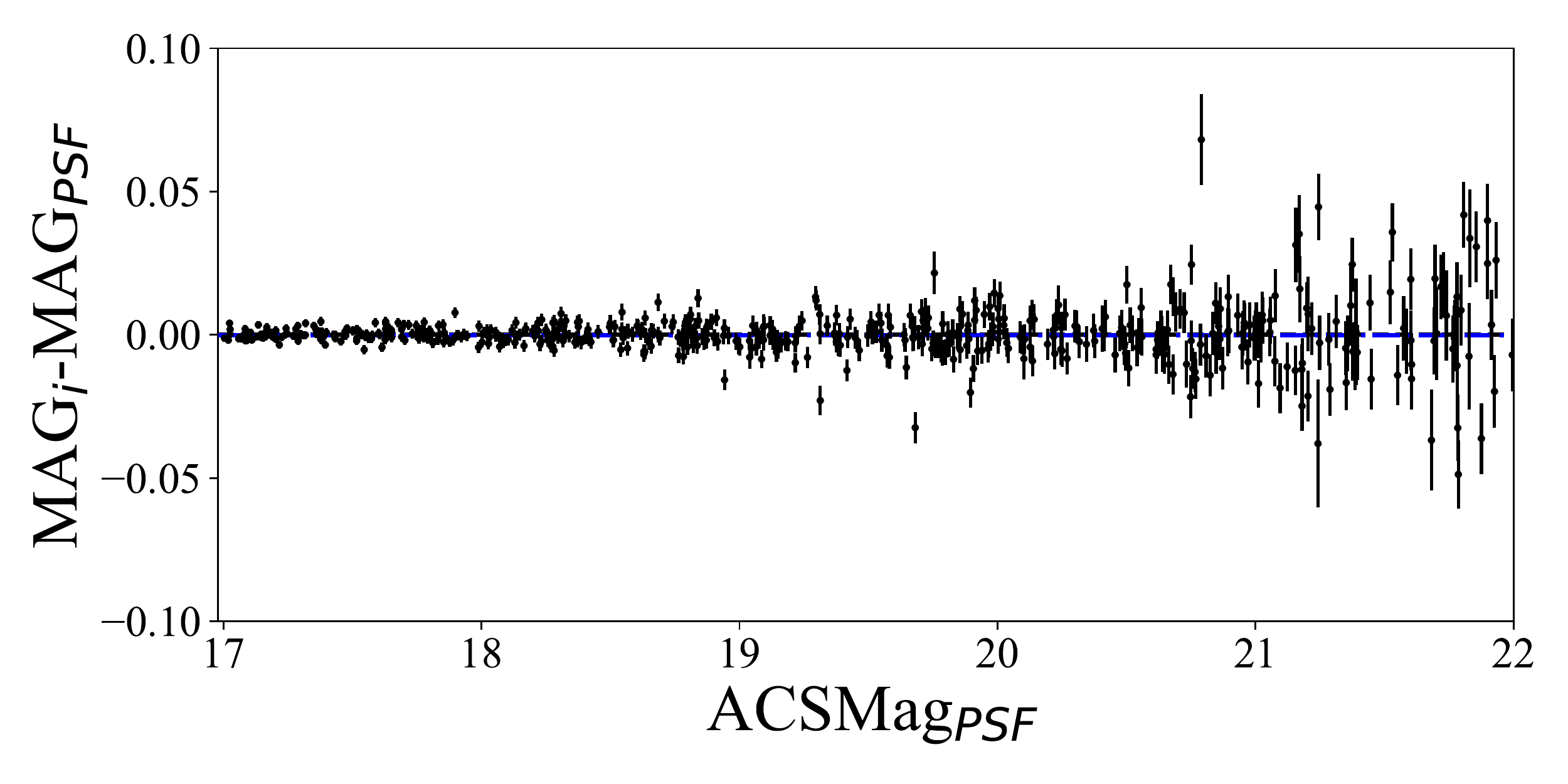}\\
    \includegraphics[width=0.31\textwidth]{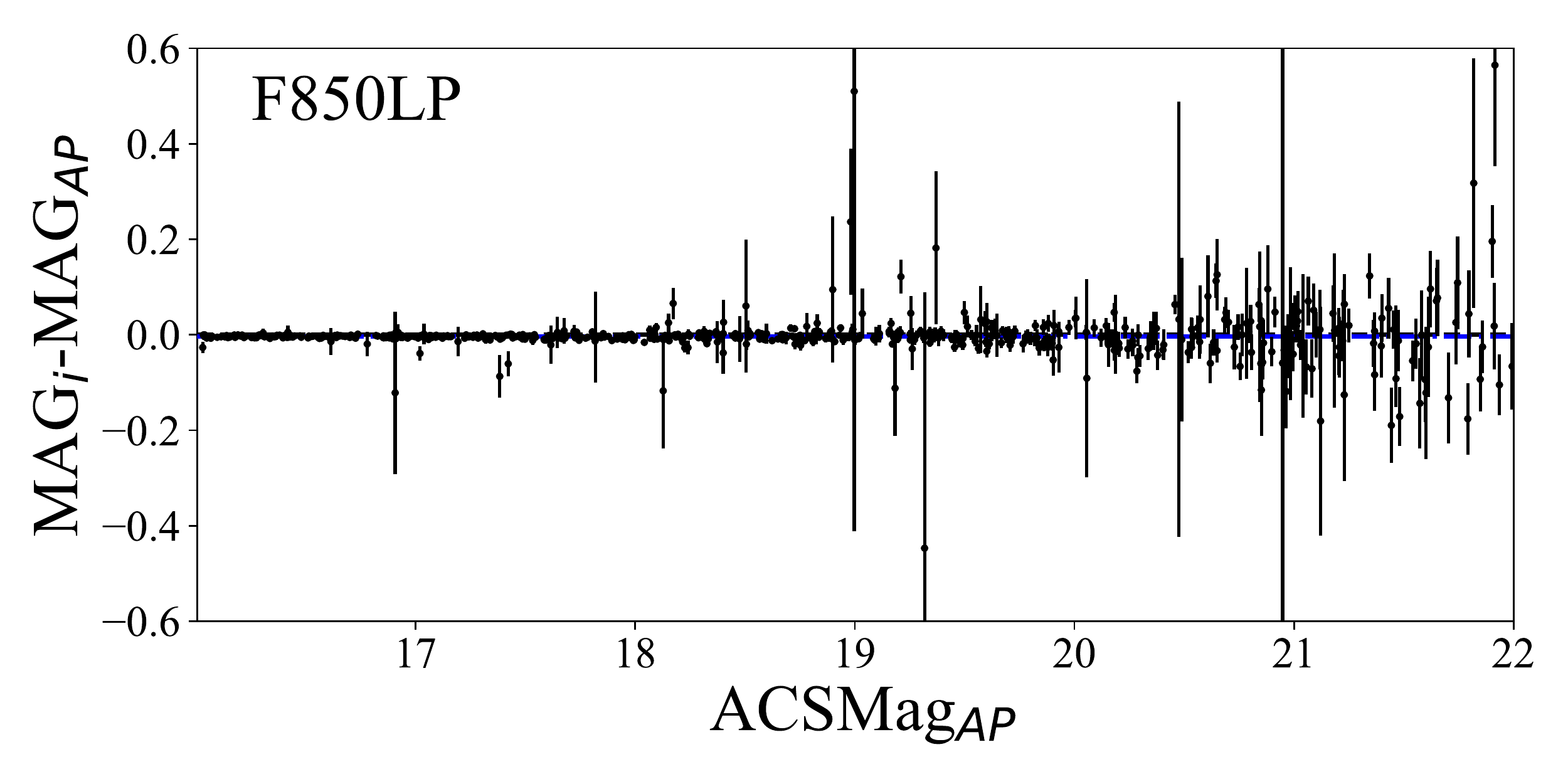}
    \includegraphics[width=0.31\textwidth]{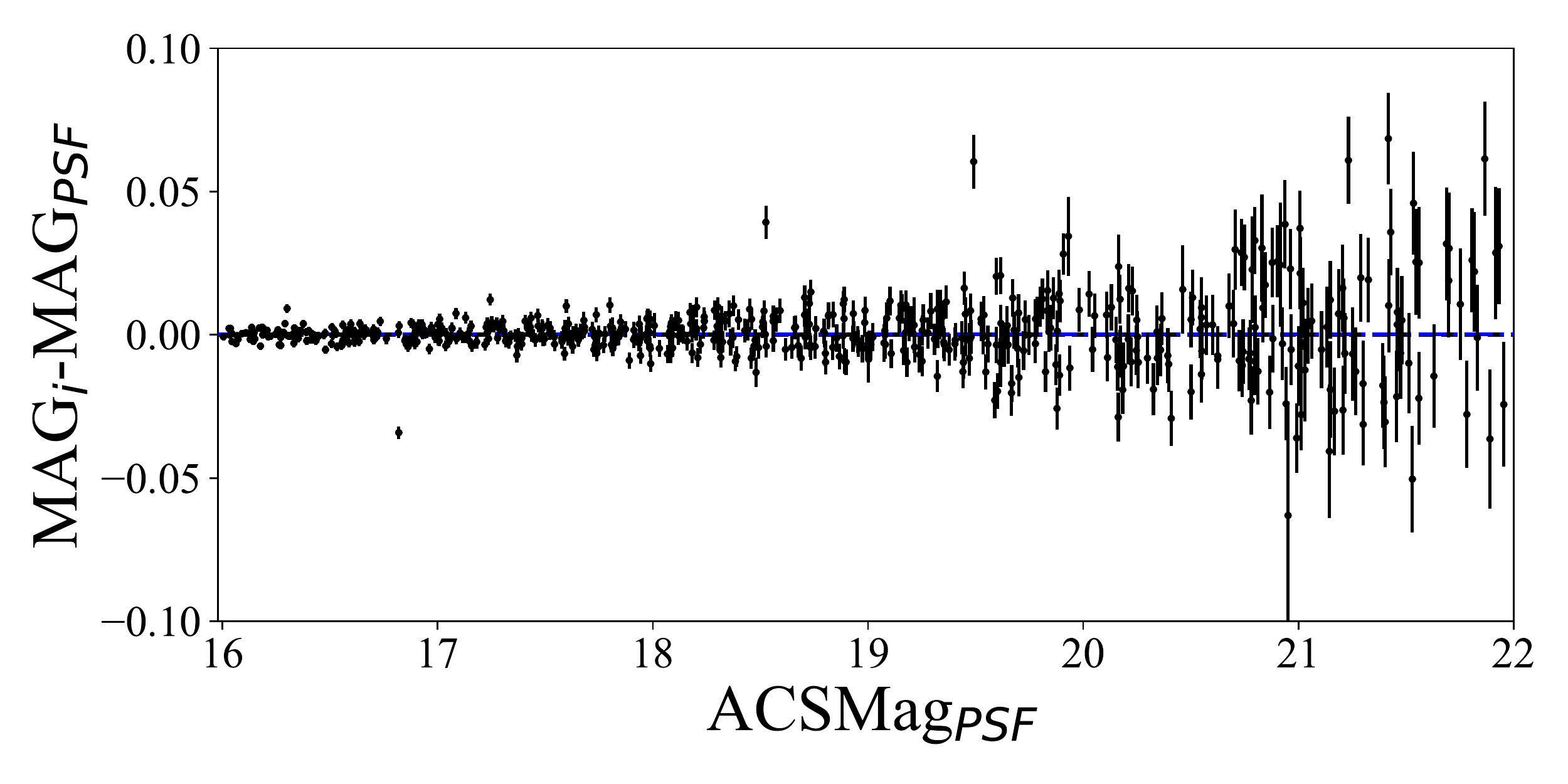}\\
    \caption{The plot shows the deltas between the known Mag$_i$ and the Mag$_x$ vs Mag$_x$, where the subscribed  Mag$_x$ represent the output of one of the photometric routines applied to isolated stars: \textit{aperture} photometry (left) and the \textit{PSF} photometry (right). Each row shows the results for each filter (F435W, F555W, F658N, F775W and F850LP) analyzed in this simulation. The constant fit is shown as a blue dash-dotted line while the black dashed line shows the locus of points where Mag$_i$=Mag$_x$. Note that the Y-axis range may differ moving from one technique to the other.}
    \label{fig:phot_deltas}
\end{figure*}

To keep these simulation as close as possible to reality, the stars have been generated covering for each filter a magnitude range matching the \cite{Robberto2013} catalog of Orion Nebula Cluster sources, constrained to prevent stars to saturate or being too faint to be detected given the exposure time. The stars have then been divided in bins with  width equal to 1 magnitude. 
Once a magnitude bin is selected, a random magnitude  is extracted (we will refer to this reference magnitude as $Mag_i$) and converted in temperature using a 1~Myr isochrone from the bt-Settl family of models. Then a random subpixel shift is generated for the artificial target and an on-sky source's coordinate is randomly selected from \cite{Robberto2013} catalogue. The coordinates are chosen so that all simulated sources are equally distributed both on sky and on the detector. 
The CCD and the coordinates of the catalog source, as well as the random pixel phase and the magnitude-dependent temperature are then used to select the closer PSF in the PSF datacubes previously generated.
This PSF is rescaled to the selected flux using the standard HST zero points  and the background is added to each pixel. This last step is performed using the \textit{make\_noise\_image}\footnote{\url{https://photutils.readthedocs.io/en/stable/api/photutils.datasets.make_noise_image.html}} from \textbf{\texttt{photutils}} that extract each background pixel value from a Gaussian distribution with mean given by the $3\sigma$-cut median of the sky evaluated at the source's real on-sky coordinates and standard deviation given by the corresponding scatter. To conclude this process, Poisson noise is added to the final tile using the \textit{apply\_poisson\_noise}\footnote{\url{https://photutils.readthedocs.io/en/stable/api/photutils.datasets.apply_poisson_noise.html}} from \textbf{\texttt{photutils}} as well.

The simulated star is then passed to the \textbf{\texttt{photometry$_{AP}$}} and \textbf{\texttt{photometry$_{PSF}$}} routines to extract photometry. Because for each position multiple PSF were generated with subpixel shifts, \textbf{\texttt{photometry$_{PSF}$}} subtracts from the target all the subpixel shifted PSFs (normalized by the total flux in the tile), choosing the one that minimizes the residual of this operation to use as reference PSF. This process has been repeated for each filter on a sample of $\sim 1500$ artificial stars. 

Figure \ref{fig:phot_deltas} and Table \ref{tab:fit_s} show, for each filter, the output of the \textit{PSF} photometry and \textit{AP} photometry routines when compared to the known input.
The discrepancies have been analyzed finding the averaged value of the three delta magnitude distributions,  i.e., Mag$_i$ - Mag$_{AP}$ vs Mag$_{AP}$, Mag$_i$ - Mag$_{MF}$ vs Mag$_{MF}$ and Mag$_i$ - Mag$_{PSF}$ vs  Mag$_{PSF}$, to find the presence of systematic shifts when compared to the known input. 

The main result is that \textit{PSF} photometry (right panel in the figure) offers the best trade-off between accuracy and precision, usually providing the smaller delta with the smaller uncertainties.

No system is perfect though, so even if the routines try to account for missing flux due to their use of a  finite aperture (\textbf{\texttt{aperture$_{AP}$}}) the limited extent of the PSF used for the fit (\textbf{\texttt{aperture$_{PSF}$}}) and the imperfect estimate of the background (despite having applied for all the tree systems the grow curve correction), cause an offset between the input flux and the measured value. Evaluating these deltas as we have done in our specific test can help mitigate this problem (see Table \ref{tab:fit_s}), providing a insight about when to correct real data or not, in cases where the input magnitudes of the sources is unknown. Comparing the results in Table \ref{tab:fit_s}, in our simulation we observe that the \textbf{\texttt{photometry$_{AP}$}} and \textbf{\texttt{photometry$_{PSF}$}} yield systematic errors that overall are smaller than the uncertainties on the \textit{deltas}. Therefore,  trying to correct for this difference in these  two cases might be counterproductive, since the correction of a small systematic delta can introduce an even larger uncertainty. 

The corrections take the form:
\begin{equation}
    m_{cor,i}=m_{i} +\Delta{i}
    \label{eq:mag_corr}
\end{equation}
where m$_{i}$ and m$_{corr,i}$ are the measured  and corrected magnitudes, $\Delta_{i}$ is the averaged shift of the measures from the expected value by the \textit{PSF/AP} \textit{photometry$_i$} tools.
The associated uncertainties can be obtained by adding the spread of the simulations to the magnitude error as:
\begin{equation}
    \begin{split}
        \label{eq:emag_corr}
        em_{corr,i}=\sqrt{em_{i}^2 +\sigma_{i}^2}
    \end{split}
\end{equation}
where $\sigma_{i}$ represent the spread of the measures at a given magnitude bin.

\subsubsection{Performance on binary systems}
\begin{figure*}[!th]
    \centering 
    \includegraphics[width=1\textwidth, angle=0]{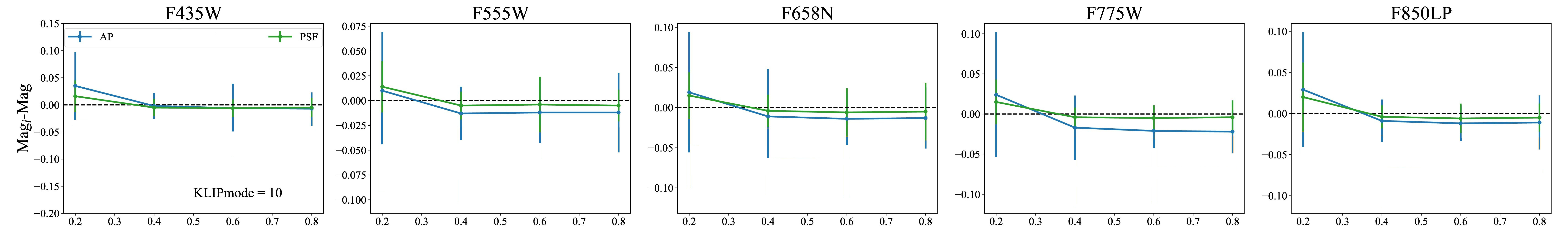}
    \includegraphics[width=1\textwidth, angle=0]{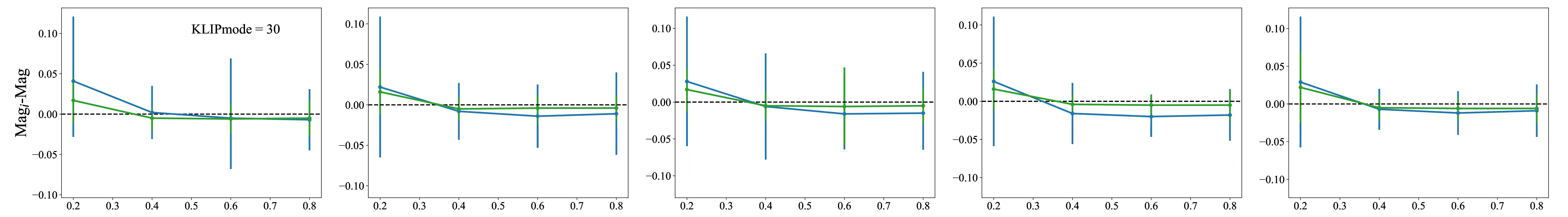}
    \includegraphics[width=1\textwidth, angle=0]{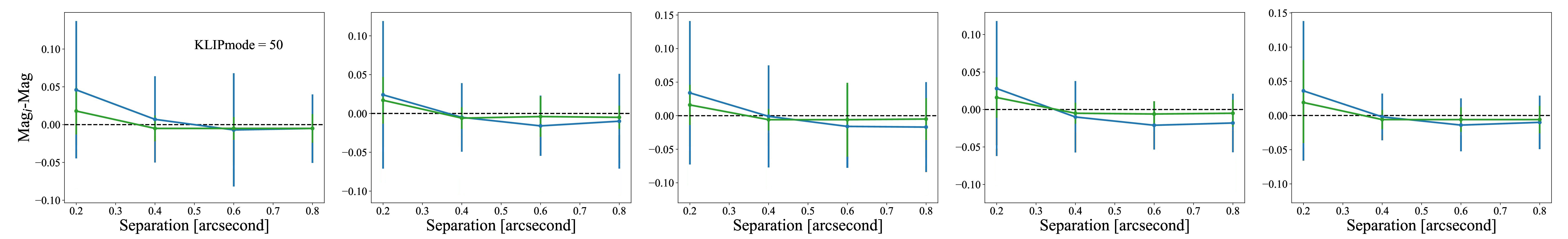}
    \caption{Results of the linear fit applied to the distribution of deltas for the primaries (Mag$_i$ - Mag$_x$) vs the separation between primary and companions, where the  Mag$_x$ represents the output of one of the photometric routines applied on the isolated primary stars. Each column represents a different filter analyzed in the survey while each row refers to a different KLIPmode. The black dashed line shows the locus of points where Mag$_i$=Mag$_x$. Note that the Y-axis range may differ moving from one to another klipmode.}
    \label{fig:prim_phot_deltas}
\end{figure*}
\begin{figure*}[!th]
    \centering 
    \includegraphics[width=1\textwidth, angle=0]{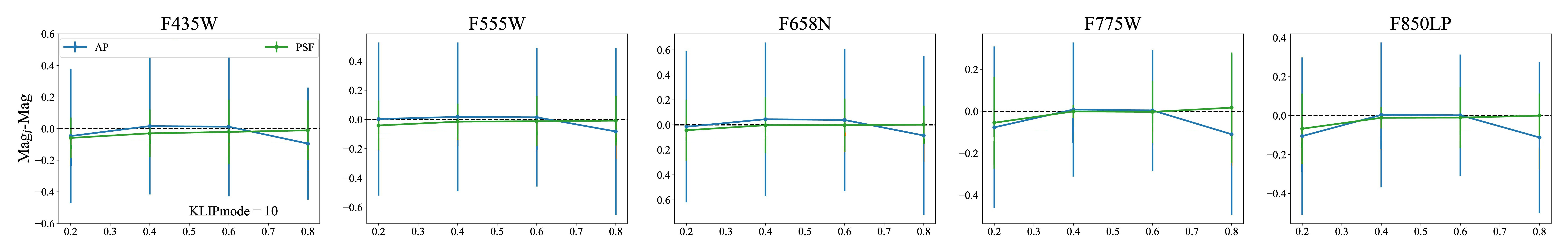}
    \includegraphics[width=1\textwidth, angle=0]{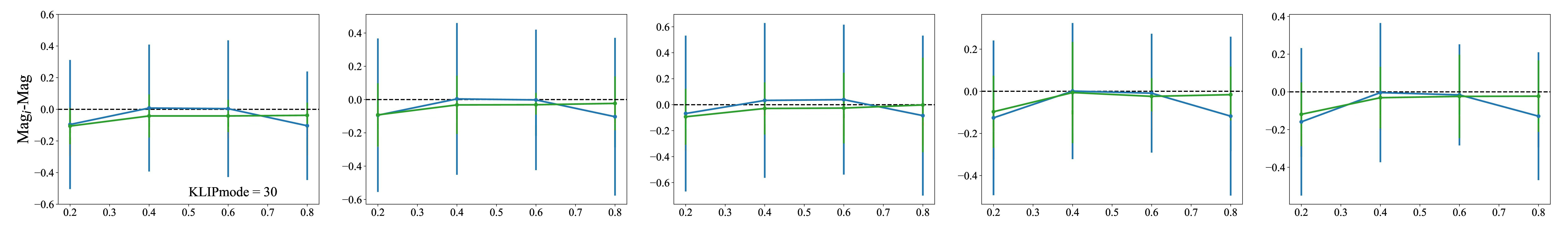}
    \includegraphics[width=1\textwidth, angle=0]{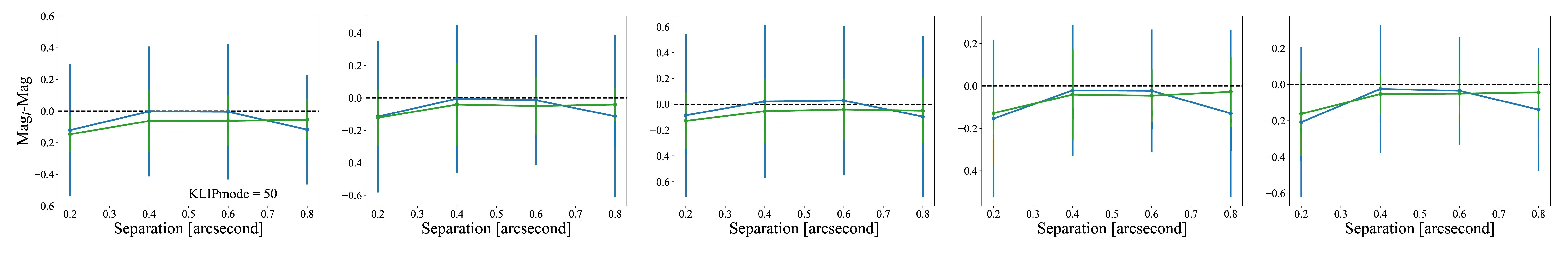}
    \caption{Similar to Figure \ref{fig:prim_phot_deltas} but for isolated companions}
    \label{fig:comp_phot_deltas}
\end{figure*}
Similar to the case of isolated star, the \textbf{\texttt{photometry$_{AP}$}} and \textbf{\texttt{photometry$_{PSF}$}} routines have been tested in the presence of binaries to understand their overall performance. As in the previous case, 1500 pairs have been simulated for each filter and for a range of separation ranging between 3 and 10 pixels, in steps of one pixel. The procedure to generate a binary is in many ways similar to the one for an isolated star: two PSFs are selected and rescaled to match the flux of the primary and companion.
Then, the companion PSF is injected in the tile of the primary at the given distance, with random orientation. Once the companion in injected, both background and Poisson noise are added as explained in section \ref{sec:Isolated stars}.

The final tile containing the binary system is then parsed to the routine that performs PSF subtraction. The \textit{residual tile} is then subtracted to obtain a tile with the isolated primary  while the \textit{residual tile} itself contains the companion. Both tiles are then parsed to the \textbf{\texttt{photometry$_{AP}$}} and \textbf{\texttt{photometry$_{PSF}$}} routines photometry and the output is compared to the original known flux of the primary and companion. Even though this is primarily a test for the photometry of the pipeline, it also yields a second relevant result: since we know a-priori the position of the injected companion, one can assess the performance of the pipeline in detecting a real companion. If the companion is detected at a position that does not coincide, within one pixel, with the coordinate of the injection, one is in presence of a false positive, and the photometry must be rejected. Analyzing the rate of success and rejections, one can understand the overall performance of pipeline to actually find companions as a function of the configuration of the binary system (i.e. filter, magnitude of the primary, difference in magnitude between companion and primary, separation, KLIPmode selected for the subtraction etc.). As discussed in the next section, this analysis can be used to estimate the rate of false positive and true positive detections as a function of the configuration of the binary, and help strengthen any decision about detecting a companion or refusing a candidate.

For the sources correctly identified, we performed a study similar to the one previously presented for isolated stars for both primaries and companions. But contrary to the case of isolated stars, the analysis of binaries depends on a much wider group of variables. One needs to consider the deltas Mag$_i$ - Mag$_x$ versus the contrast (i.e. the difference in magnitude between companion and primary), where the  Mag$_x$ represent the output of one of the photometric routines applied on the isolated primary stars itself. Secondly, each distributions must be evaluated for each filter, for each separation and for the KLIPmode utilized to truncate the Karhunen-Lo{\`e}ve transformation. We summarize the results in Figures \ref{fig:prim_phot_deltas} and \ref{fig:comp_phot_deltas}, showing the outcome of the constant fit applied on each distributions as a function of the separations, with error bars representing the relative uncertainty on each fit. Each columns refer to a different filter while each row refer to a different KLIPmode.
 
As expected, when a close companion (separation $\leq 0.2''$) is subtracted from a binary the outcome of the \textbf{\texttt{photometry$_{AP}$}} and \textbf{\texttt{photometry$_{PSF}$}} routines is worst than for wider binaries (separation $> 0.2''$) due to the increasing overlap between the two PSFs. 

\begin{figure*}[ht!]
\begin{center}
\includegraphics[width=0.75\textwidth]{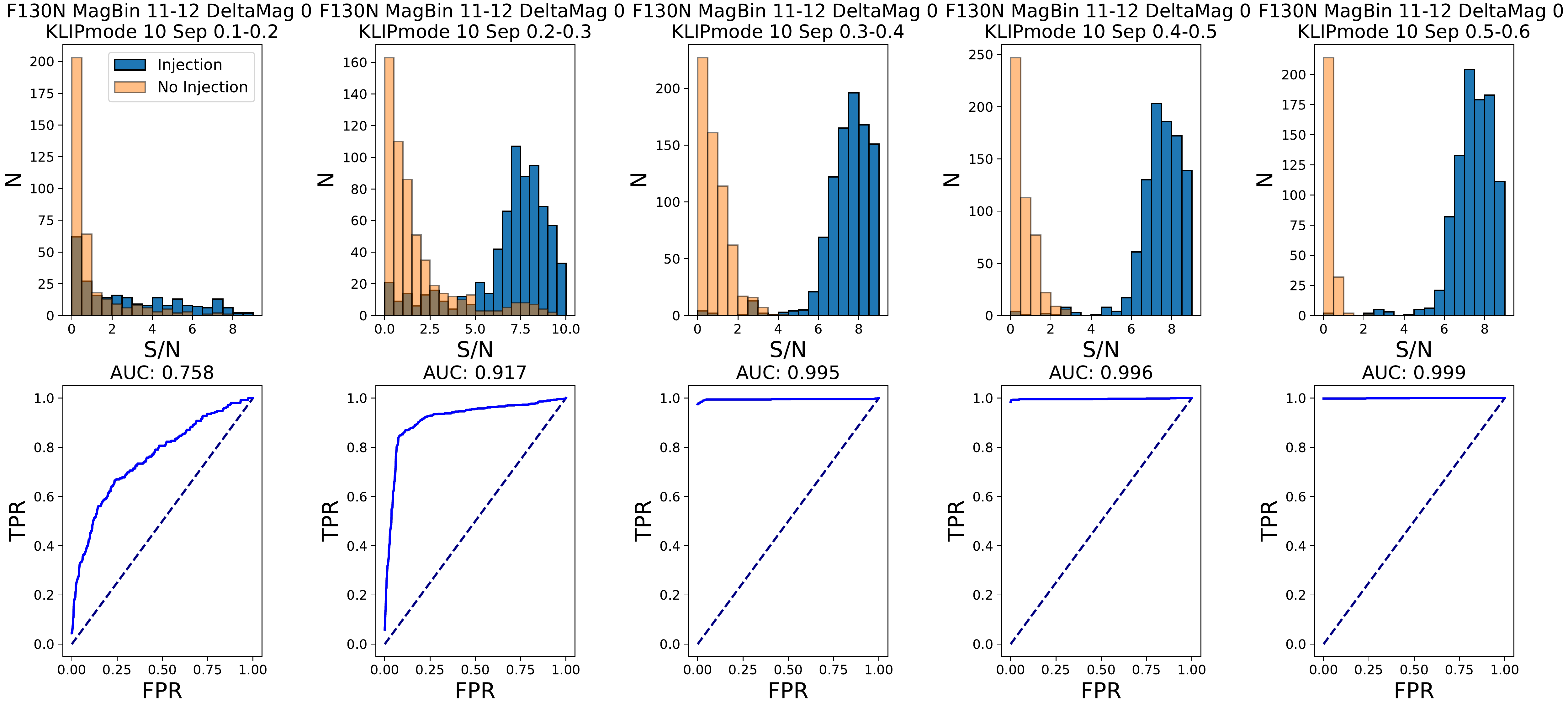}\\
\includegraphics[width=0.75\textwidth]{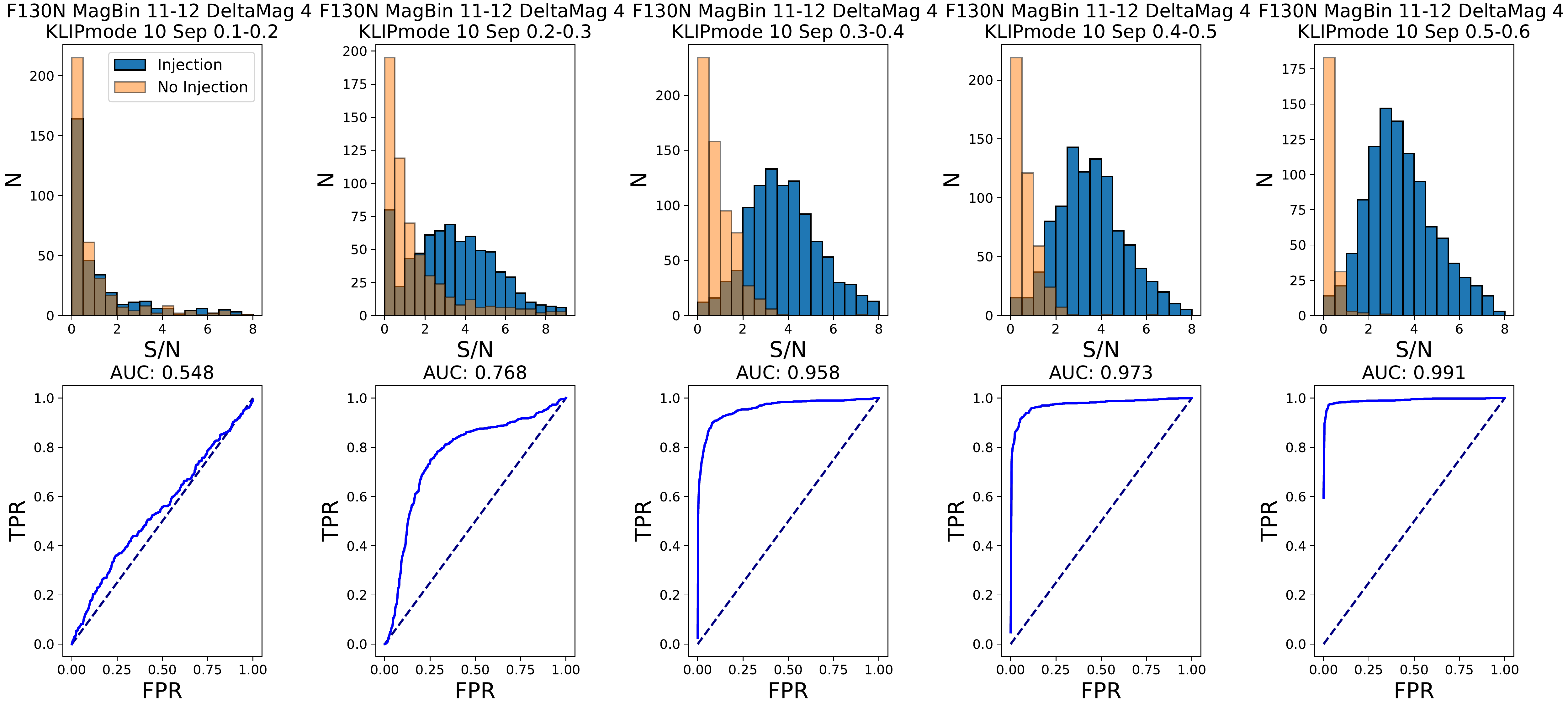}\\
\includegraphics[width=0.75\textwidth]{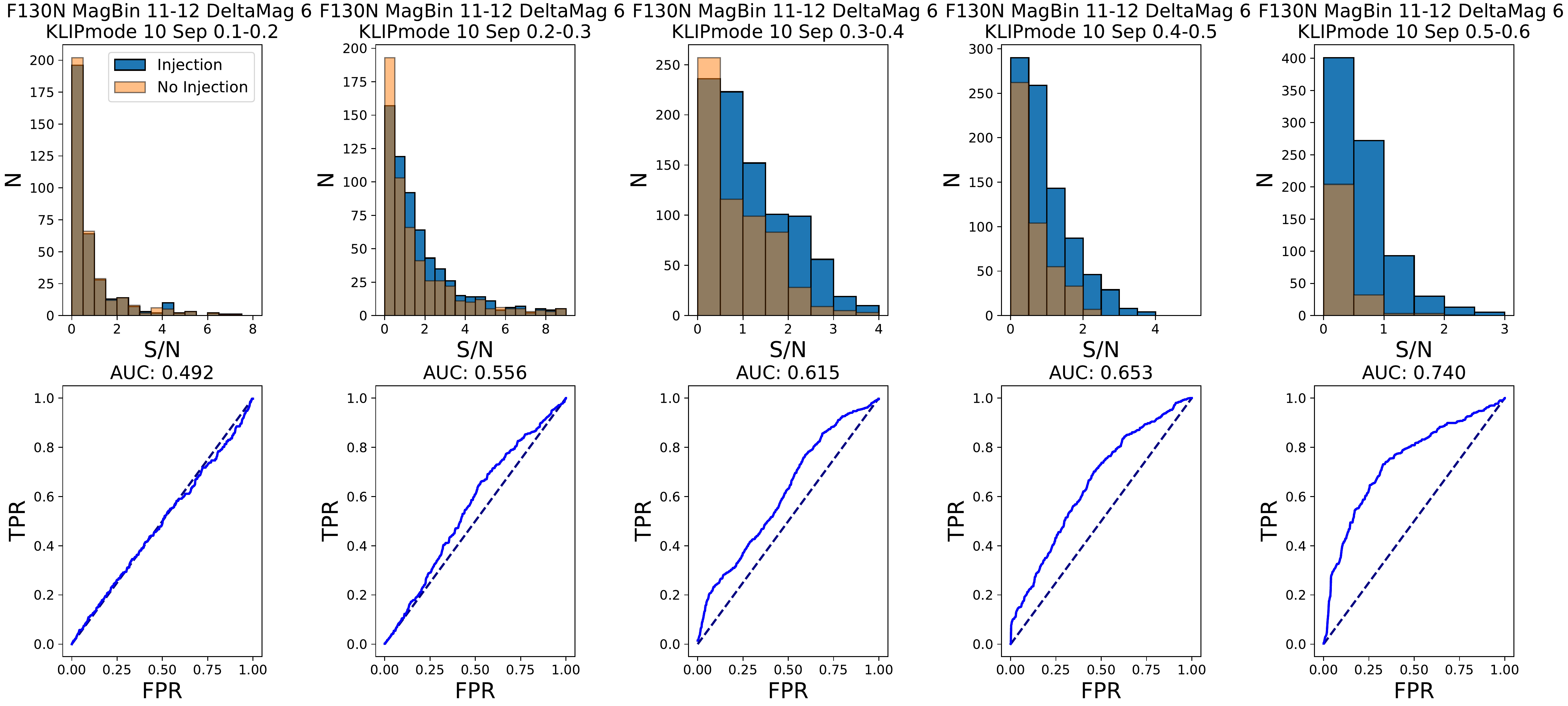}\\
\caption{Distributions of signal-to-noise  and derived ROC curves for filters F130N, magnitude bin of the primary 11-12, $\Delta\mathrm{mag}$ 0,4 and 6 and different distances from the center of the tile. From: \cite{Strampelli2020} \label{Fig:FINJ_11_0}}
\end{center}
\end{figure*}
For wide binaries, similar to the case of isolated stars, the \textbf{\texttt{photometry$_{PSF}$}} appear more precise compared to the \textbf{\texttt{photometry$_{AP}$}}. However, both \textbf{\texttt{photometry$_{AP}$}} and \textbf{\texttt{photometry$_{PSF}$}} show problems in correctly evaluate the photometry of an isolated primary when a companion is found very close to the hosting star (separation $\leq 0.2''$) due to an uncompleted subtraction of the close-in companion, that in turn brings to an overestimate of the isolated primary. This trend is reflected in the isolated companions where both routines tend to underestimate the flux of close-in companion, bringing to an underestimate of the isolated companion. In this case, smaller KLIPmode are able to achieve better results than bigger KLIPmode, when dealing with close-in companions

Moreover, due to the finite dimension of the tiles we used in the simulation, when a companion is very close to the border of the tile (separation $\gtrsim 0.8''$), \textbf{\texttt{photometry$_{AP}$}} fairly fails to recover the correct flux of the isolated companions, while the \textbf{\texttt{photometry$_{PSF}$}} is able to correctly reconstruct the flux of the isolated companion. 

\section{\textbf{STEP 9: False positive analysis}}
\label{sec:False Positive Analysis}
In this section we present  the infrastructure developed to characterize the detection reliability. 
This final step relies heavily on the simulation of isolated stars and binaries described in Section \ref{sec:Pipeline photometry performance}, where we have shown how to inject artificial companions and isolated sources. 

Once a library of both singles and binaries has been created, it can be passed to the pipeline for analysis. It is important to notice that each binary system is linked to a corresponding single source tile by the same primary star, i.e. the tiles are created in tandem: one with a single source, and one with the same source plus an injected companion with random flux and random position \textit{i,j}. In this way the pipeline can measure the flux at the injected position \textit{i,j} both for the isolated source (building the null hypothesis) and for the linked binary (building the test hypothesis). As shown in \cite{Strampelli2020}, one can use Receiver Operating Characteristic \textbf{\citep[ROC; ][]{Fawcett2006}} curve as a tool to assess the ability of a binary classification system; the discrimination threshold \textit{T} provides an estimate of the probability to have false positives as a function of the particular configuration of the binary (i.e. filter under examination, the magnitude of the primary, delta magnitude between companion and primary, KLIPmode utilized to detect the companion and separation between the two).


The pipeline can build ROC curves and provide the TPR and FPR population representative of each candidate. These curves show how the achieved sensitivity strongly depends on the configuration of the binary: the magnitude of the primary ($\mathrm{m_{F130M}}$), the contrast ($\Delta\mathrm{mag}$) achieved by PSF subtraction, and the distance of the companion from the primary (separation) and the KLIPmode utilized during the PSF subtraction. 

To encapsulate in a single number the performance of our model to distinguish between classifier, the pipeline will evaluate the Area Under the Curve (AUC) of an ROC. The higher the AUC, the better is the model at distinguishing between the true positive population and the false positive population. An AUC of 1 indicates a perfect classifer.

Figure \ref{Fig:FINJ_11_0} shows examples of the TP (blue) and FP (orange) histograms for a given binary configurationf, and the corresponding ROC curve. Also provided for each ROC curve is the value of the corresponding AUC. 
For a specific application of this methodology to WFC3-IR data, see \cite{Strampelli2020}.

\begin{figure*}[!th]
    \centering
    \includegraphics[width=0.95\textwidth]{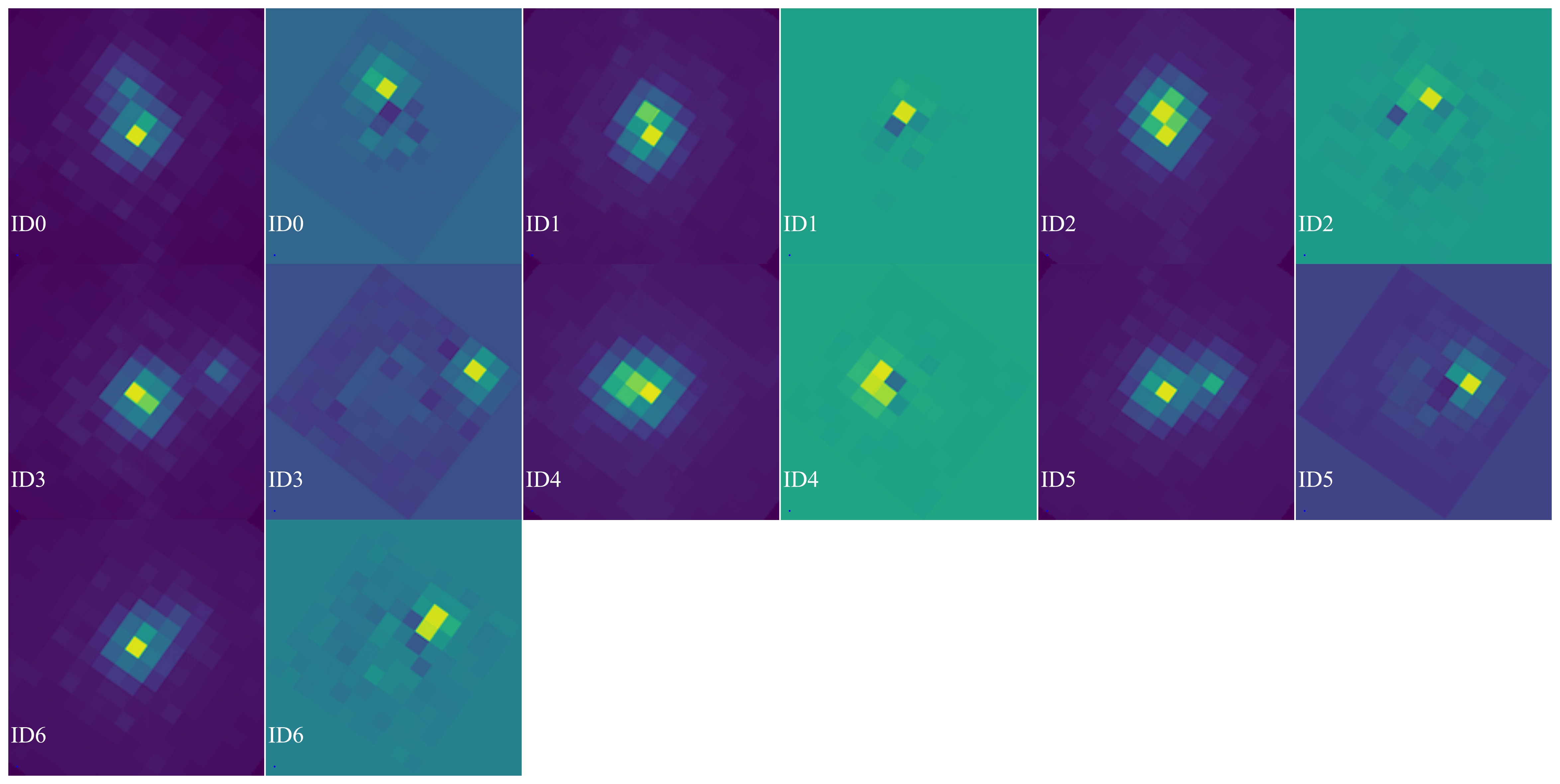}\\
    \includegraphics[width=0.95\textwidth]{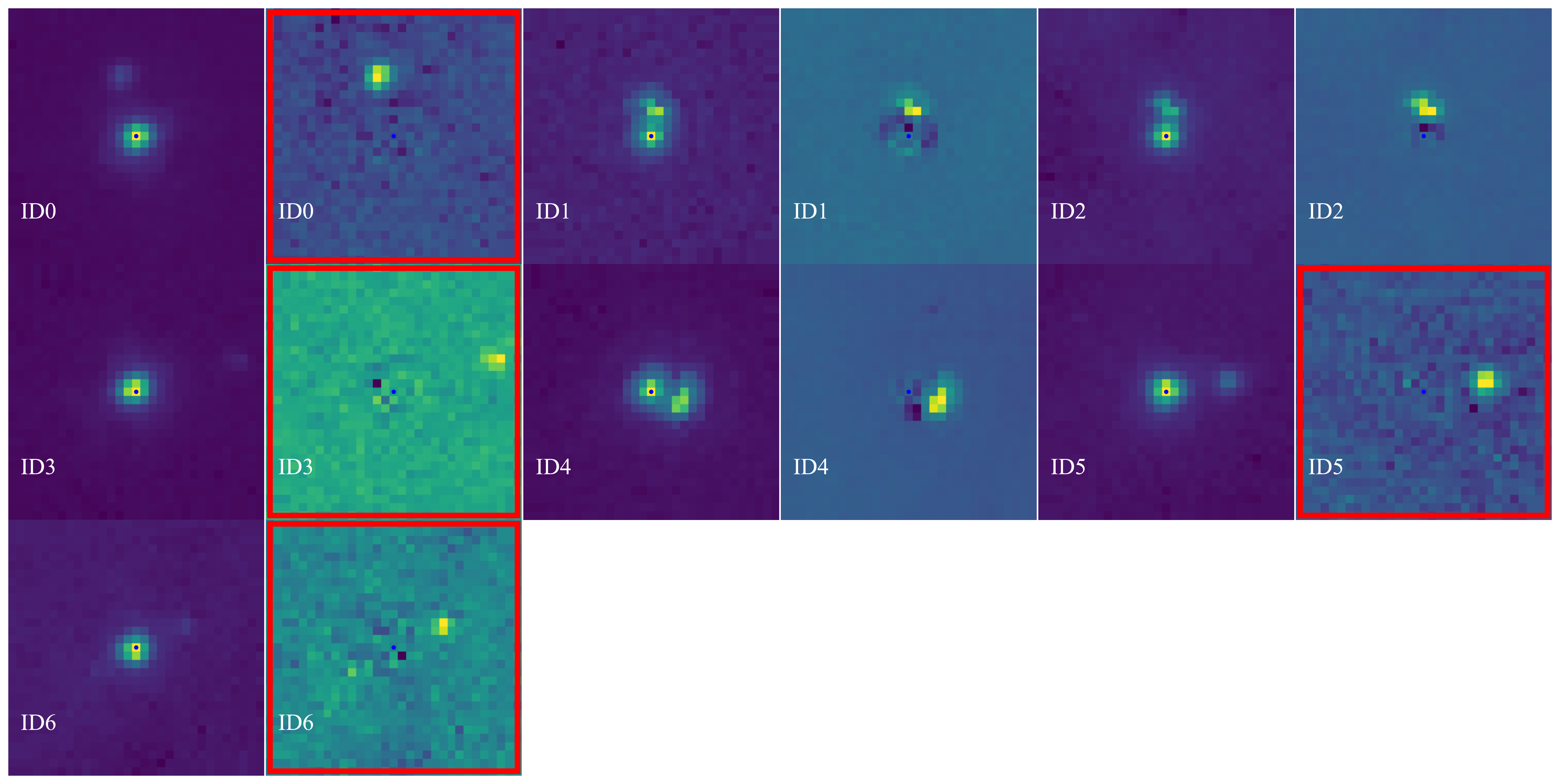}\\
    \caption{Tiles before/after KLIP PSF subtraction for a sample of targets from WFC3-IR data (upper 2 rows) and correspondent detection in ACS data (lower 2 rows). For each ID, the first tile represent the system before KLIP, while the second shows the residual tile with the detected companion. A red square marks the residual tiles for binaries already known in the input catalog (shown here for comparison with KLIP detections) . Each tiles has a dimension of 2''x2''. The north is up and east is on the left. From: \cite{Strampelli2020,Strampelli2022}}
    \label{fig:WFC3-ACS comparison}
\end{figure*}

\begin{figure*}[!th]
    \centering
    \includegraphics[width=0.45\textwidth]{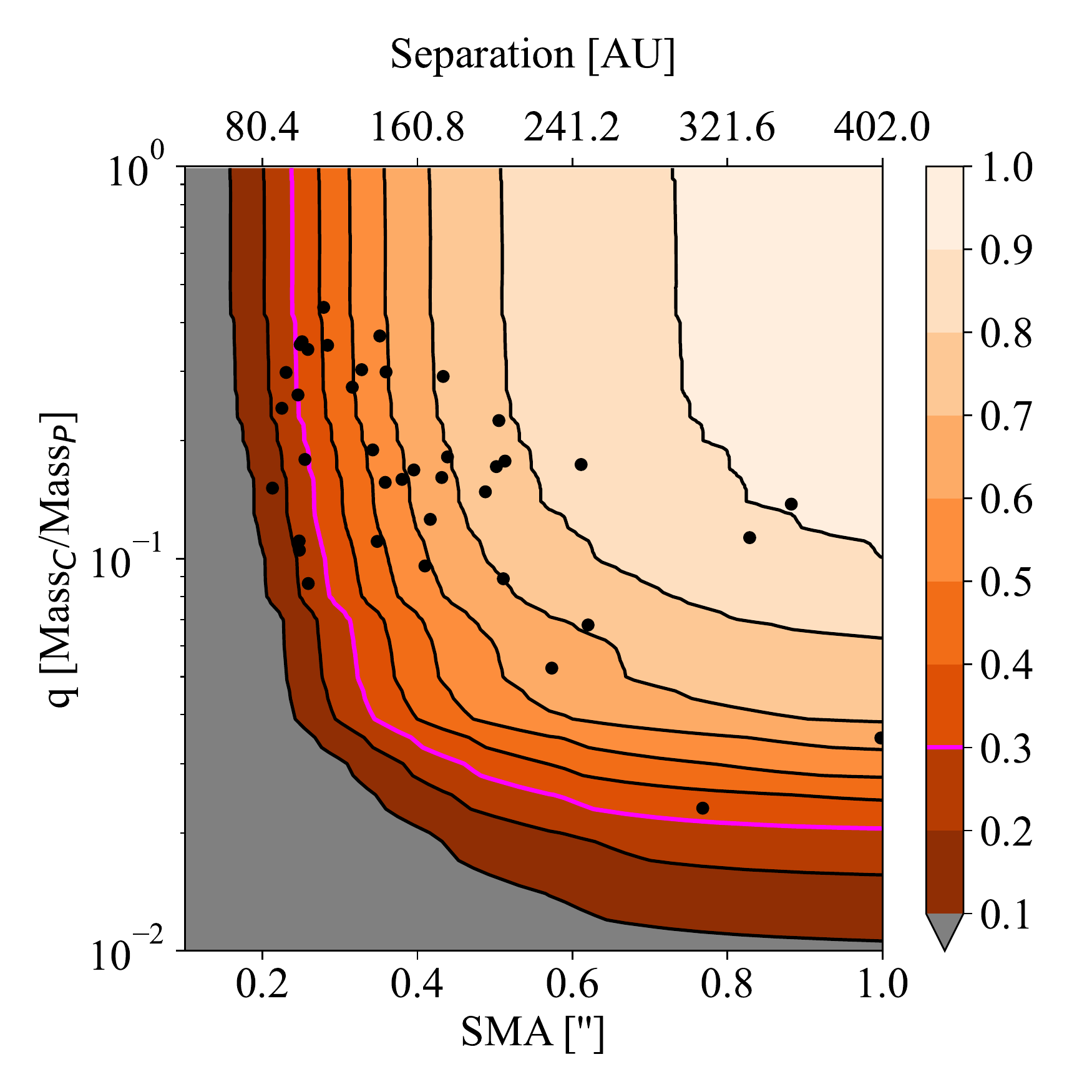}
    \includegraphics[width=0.44\textwidth]{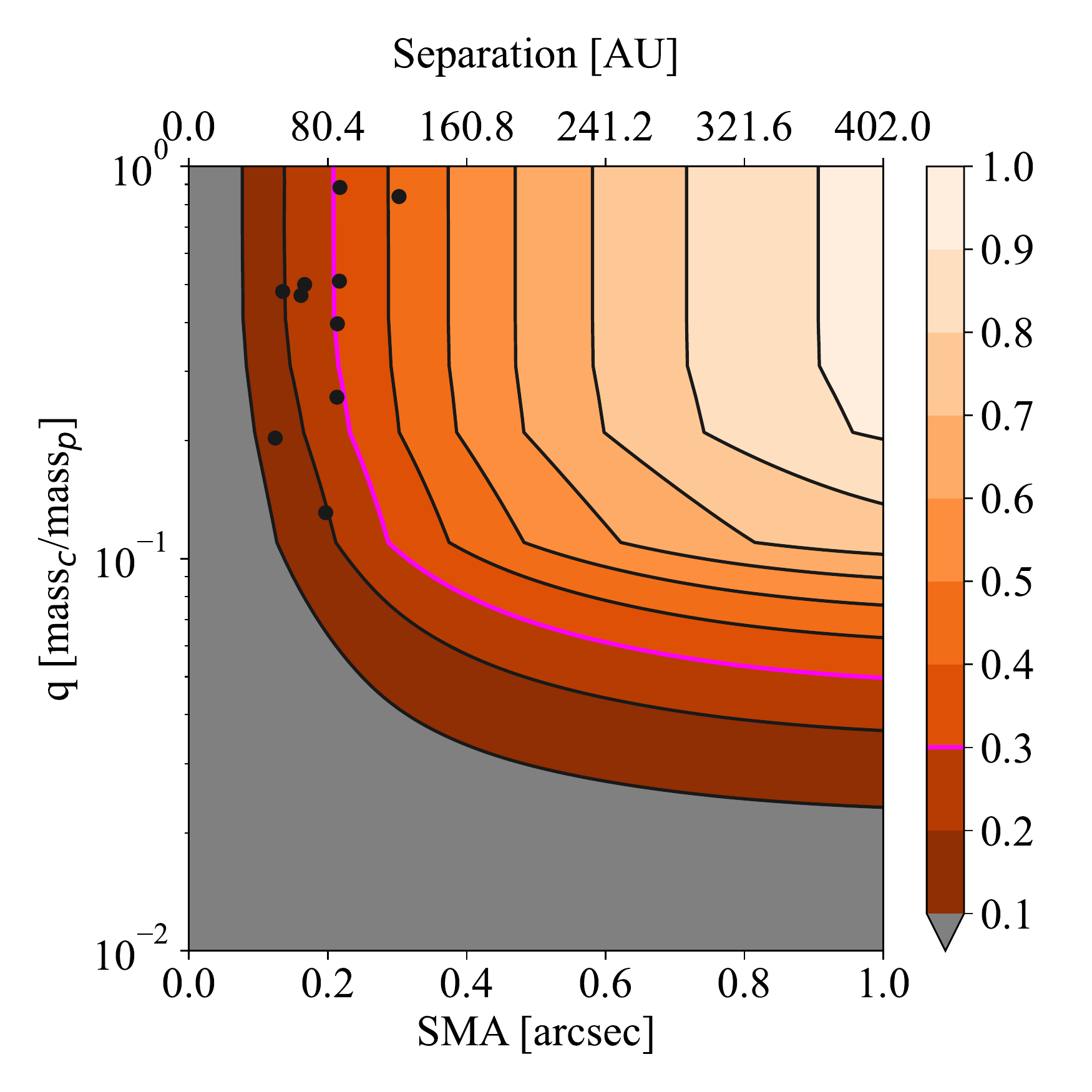}
    \caption{Mass ratio \textit{q} completeness function of projected SMA for WFC3-IR data (left) ad ACS data(right). The black dots marks the position of each candidate detected in that work, while the magenta line mark the $30\%$ completeness threshold for both instruments. From: \cite{Strampelli2020,Strampelli2022}}
    \label{fig:q_completeness}
\end{figure*}

\begin{figure}[!th]
    \centering
    \includegraphics[width=0.45\textwidth]{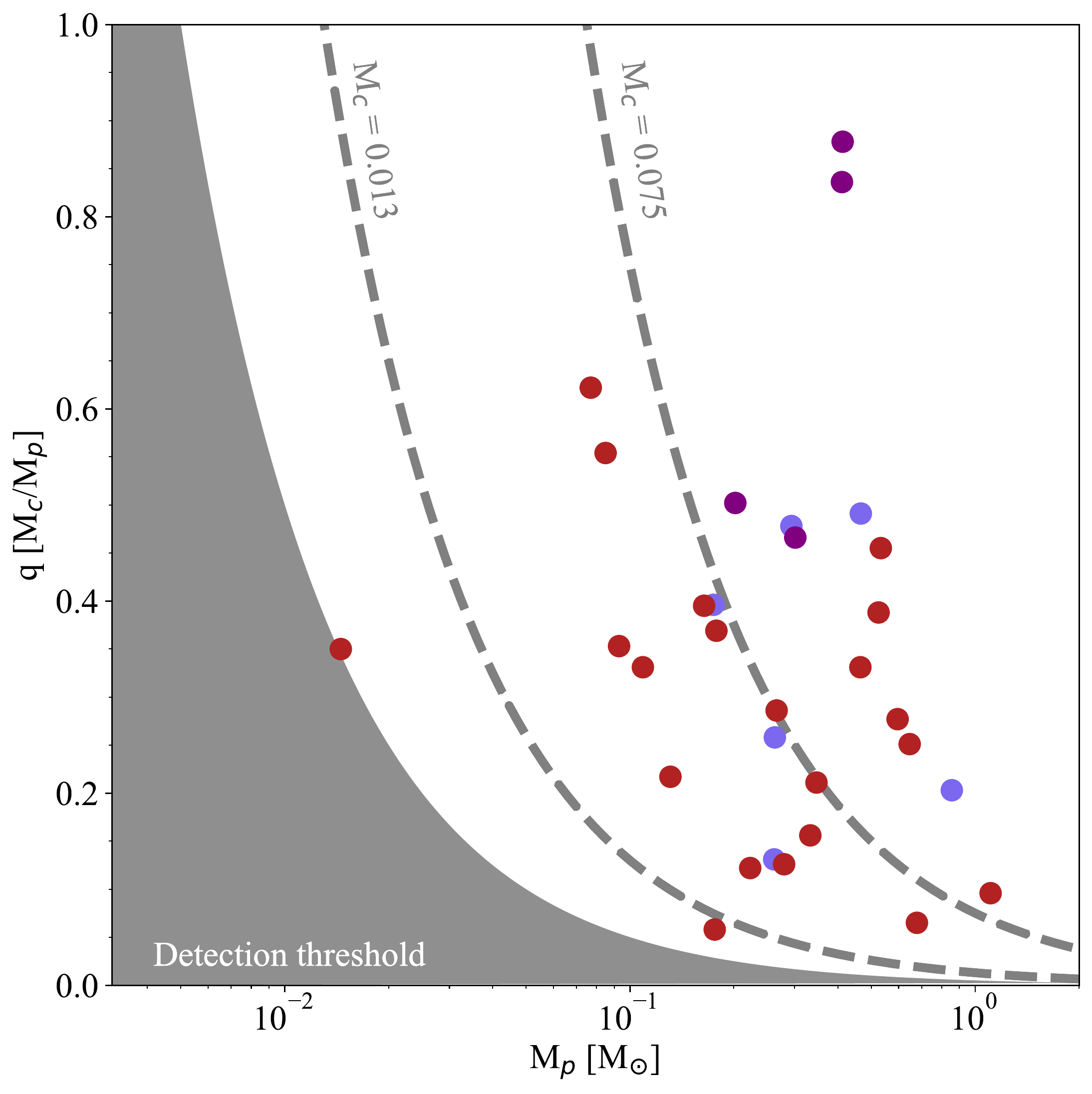}
    \caption{KLIP detected binaries mass ratio distribution as a function of the primary mass objects for the joined catalogs from the two surveys. The sources are color-coded by original instrument that provide the detection (purple: both, blue: ACS, red: WFC3-IR).}
    \label{fig:Qplots}
\end{figure}

We need to stress that this step does not provide a direct false positive rejection, as it only focuses on the simulated sources and knows nothing about the real ones: it does not separate single false positives from true detections in the sample, but only provide an estimate on how probable is that a specific configuration can produce false positive in the simulations. In other words, its only scope is to associate to each configuration of a binary system a probability to be the outcome of a false positive detection, that then the user can use in order to make an informed decision if a companion is plausibly a false positive or is probably real. A series of ancillary routines has been developed to perform this analysis, as shown in the two scientific application of the pipeline. For a practical application of this analysis, we highly encourage the reader to look at \cite{Strampelli2020} where this approach is deeply discussed on \textit{HST}/WFC3-IR data. 

\section{Test with real data}
\label{sec:Test with real data}
In \cite{Strampelli2020} we analyzed the HST/WFC3-IR data for the HST Treasury Program on the Orion Nebula Cluster (GO-10246) while in \cite{Strampelli2022} we performed a similar analysis focused instead on the HST/ACS data obtained for the HST Treasury Program on the same cluster (GO-10246). Here in the following we will shortly summarize some of the results from those papers to show the performance of our pipeline, but we still refer the reader to the specific papers to obtain more details.

Figure \ref{fig:WFC3-ACS comparison} shows the comparison between 7 detections from the WFC3-IR data and corresponding detection in the ACS survey (either because they are already present in the input catalog or because they were found by PSF subtraction). Note that due to the different filters employed in the two surveys, sometimes the primary may vary, or the companion may not be visible at all. Overall we find very good agreement between the detections in the two surveys.

\subsection{Completeness and multiplicity fraction}
Figure \ref{fig:q_completeness} shows the comparison between the completeness achieved in the WFC3-IR survey (on the left) and in the ACS survey (on the right). As expected, the infrared survey allows us to detect lower-mass companions, reaching a $q \sim 0.025$ at $\sim 30\%$ of completeness and separations of $\sim 100$ AU (or semi-major axis - SMA - $\gtrsim 0.8''$), while the ACS survey allows only to push less deeper in contrast and reach only $q\sim 0.05$ for similar ranges. On the other hand though, the ACS surveys allow us to go deeper inside the system and closer to the primary, achieving completeness $\gtrsim 30\%$ already at $\sim 80$ AU (or SMA $\sim 0.2''$), compared to the $\sim 100$ AU (or SMA $\sim 0.25''$) for WFC3-IR, and a minimum completeness of $\sim 10\%$ at $\sim 40$ AU (or SMA $\sim 0.1''$), compared to the $\sim 60$ AU (or $0.15''$ SMA) for WFC3-IR.

Figure \ref{fig:Qplots} instead shows the mass ratio distribution for our candidates as a function of the mass of the primary. Overall, in the specific case of the ONC, we are able to detect brown dwarf mass companions down to the limit of planetary mass and three planetary mass companions, reaching a minimum separation of $\sim 70 AU$ ($\sim 0.17''$ SMA) at a completeness of $\sim 22\%$. 

Both in \cite{Strampelli2020,Strampelli2022} we combined the observed companion population for the WFC3-IR and ACS instrument with the corresponding ROC curves and false positive analysis to obtain an unbiased multiplicity fraction in the ONC. The two works reported remarkably close results:

\begin{equation}
    \begin{split}
        MF_{WFC3-IR} = 11.5\% \pm 0.9\%\\
        MF_{ACS} = 11.9\% \pm 0.5\%
    \end{split}
\end{equation}

\section{Conclusion}
A new pipeline has been developed to detect and characterize faint astronomical sources close to their host star. It relies on Karhunen-Lo{\`e}ve truncated transformation theory to perform PSF subtraction. 
Using a highly adaptable series of routines, the pipeline uses standard imaging data and conventional photometry to deliver a catalog of candidate close companions, their photometry with associated uncertainties, and a robust statistical estimate of their false positive probability.
At the current stage of the development \textbf{\texttt{\textit{Stra}KLIP}} is  able to detect and characterize binaries. It cannot be directly applied If more than two sources are present in the same search area $\mathcal{S}$, as in the case of e.g. hierarchical triple systems. The extension to the case of multiple sources will be the subject a future version.

Even though this pipeline is based on methods designed for observations of a single star at a time, our we can achieve remarkably results on mosaic space based data. In fact, in the case of the ONC, the pipeline is able to reliably detect very low-mass companion, down into the brown dwarf mass and almost to the planetary mass limit. The pipeline is also able to detect signal as close as $\gtrsim 0.1 ''$ (or $\sim 40 AU$  at the distance of the ONC) with a completeness $\gtrsim 10\%$, or $\sim 0.2''$ ($\sim 80 AU$) with a completeness of $\sim 30\%$ . This approach can potentially be applied to a wide variety of space based imaging survey, starting with the exiting HST archive, to near future JWST mosaics, to future wide field Roman images.

\acknowledgments
GMS wants to thanks the Instituto de Astrofísica de Canarias for hospitality.  The authors thank Carlo Manara for the excellent insights. Support for Program number GO-10246 and GO-13826 was provided by NASA through a grant from the Space Telescope Science Institute, which is operated by the Association of Universities for Research in Astronomy, Incorporated, under NASA contract NASS-26555. CFM acknowledges an ESO fellowship. JA was supported in part by a grant from the National Physical Science Consortium. GMS and AA are supported by the Ministerio de Ciencia, Innovación y Universidades of Spain (grant AYA2017-89841-P) and by the Instituto de Astrofísica de Canarias.


\software{AstroSCRAPPY \citep{McCully2018}, DOLPHOT \citep{Dolphin2000}, PyKLIP \citep{Wang2015}, Pandas \citep{mckinney-proc-scipy-2010}, Photutils \citep{Bradley2020}}

\FloatBarrier

\clearpage
\bibliography{main} 

\end{document}